\documentclass[twocolumn]{aastex63}
\usepackage{gensymb}
\usepackage{hhline}
\usepackage{import}
\usepackage{amsmath}
\usepackage{dirtytalk}
\usepackage{listings}
\usepackage[nameinlink]{cleveref}

\usepackage[T1]{fontenc}

\accepted{to AJ}

\shorttitle{Eccentricity Distributions of Imaged Planets}

\graphicspath{{./}{}}

\usepackage{xcolor}

\definecolor{codegreen}{rgb}{0,0.6,0}
\definecolor{codegray}{rgb}{0.5,0.5,0.5}
\definecolor{codepurple}{rgb}{0.58,0,0.82}
\definecolor{backcolour}{rgb}{0.95,0.95,0.92}

\lstdefinestyle{mystyle}{
    backgroundcolor=\color{backcolour},   
    commentstyle=\color{codegreen},
    keywordstyle=\color{magenta},
    numberstyle=\tiny\color{codegray},
    stringstyle=\color{codepurple},
    basicstyle=\ttfamily\footnotesize,
    breakatwhitespace=false,         
    breaklines=true,                 
    captionpos=b,                    
    keepspaces=true,                 
    numbers=left,                    
    numbersep=5pt,                  
    showspaces=false,                
    showstringspaces=false,
    showtabs=false,                  
    tabsize=2
}

\begin{document}

\title{The Impact of Bayesian Hyperpriors on the Population-Level Eccentricity Distribution of Imaged Planets}

\correspondingauthor{Vighnesh Nagpal}
\email{vighneshnagpal@berkeley.edu}

\author[0000-0001-5909-4433]{Vighnesh Nagpal}
\affiliation{Astronomy Department, University of California, Berkeley, CA 94720, USA}

\author[0000-0002-3199-2888]{Sarah Blunt}
\affiliation{Department of Astronomy, California Institute of Technology, Pasadena, CA, USA}

\author[0000-0003-2649-2288]{Brendan P. Bowler}
\affiliation{Department of Astronomy, The University of Texas at Austin, TX 78712, USA}

\author[0000-0001-9823-1445]{Trent J. Dupuy}
\affiliation{Institute for Astronomy, University of Edinburgh, Royal Observatory, Blackford Hill, Edinburgh, EH9 3HJ, UK}

\author[0000-0001-6975-9056]{Eric L. Nielsen}
\affiliation{Department of Astronomy, New Mexico State University, P.O. Box 30001, MSC 4500, Las Cruces, NM 88003, USA}

\author[0000-0003-0774-6502]{Jason J. Wang}
\affiliation{Department of Astronomy, California Institute of Technology, Pasadena, CA, USA}

\begin{abstract}

Orbital eccentricities directly trace the formation mechanisms and dynamical histories of substellar companions. Here, we study the effect of hyperpriors on the population-level eccentricity distributions inferred for the sample of directly imaged substellar companions (brown dwarfs and cold Jupiters) from hierarchical Bayesian modeling (HBM). We find that the choice of hyperprior can have a significant impact on the population-level eccentricity distribution inferred for imaged companions, an effect that becomes more important as the sample size and orbital coverage decrease to values that mirror the existing sample. We reanalyse the current observational sample of imaged giant planets in the 5-100 AU range from \citet{bowler:2020} and find that the underlying eccentricity distribution implied by the imaged planet sample is broadly consistent with the eccentricity distribution for close-in exoplanets detected using radial velocities. Furthermore, our analysis supports the conclusion from that study that long-period giant planets and brown dwarf eccentricity distributions differ by showing that it is robust to the choice of hyperprior. We release our HBM and forward modeling code in an open-source Python package, \texttt{ePop!}, and make it freely available to the community. 

\end{abstract}

\keywords{planets and satellites: fundamental parameters, planets and satellites: dynamical evolution and stability, stars: brown dwarfs}

\section{Introduction} 

The orbital eccentricities of exoplanets reflect the physical processes that sculpt the formation and dynamical evolution of planetary systems. Models of giant planet formation from the axisymmetric accretion of gas within a protoplanetary disk lead to circular, coplanar orbits \citep{armitagebook}. Subsequent interactions with other planets (e.g.,  \citealt{rasio_ford}; \citealt{juric_tremaine}; \citealt{dawson_murrayclay}), the disk itself (e.g., \citealt{goldreich_sari}), or massive outer companions (e.g., \citealt{naoz}) can increase a planet's eccentricity. Once these eccentricities have been excited, however, dissipatory forces (e.g., \citealt{ogilvie}) and torques from the disk and central star can act to damp orbital eccentricities over time (e.g., \citealt{chiang_duffell}; \citealt{morbidelli}). The statistical properties of planet eccentricities at different ages, planet masses, orbital separations, and host star masses will help unravel the dominant physical processes at play in planet formation and evolution, which can be challenging to infer in individual systems.

The dominant theories of giant planet formation at wide separations predict qualitatively different population-level eccentricity distributions. Unperturbed objects formed within disks through core (or pebble-assisted) accretion or disk instability are expected to have low eccentricities. Conversely, companions that form through cloud fragmentation or experience outward scattering are expected to have eccentricities spanning a wide range (e.g.,  \citealt{veras}, \citealt{bate}). 

The growing population of exoplanets discovered using direct imaging presents an exciting opportunity to test planet formation mechanisms. These self-luminous giant planets are primarily young and orbit over a wide range of separations (5-10000 AU; \citealt{bowlerreview}). The inner-most planets from this larger sample ($\sim$ 5-100 AU) have detectable orbital motion, which enables their eccentricities to be constrained (e.g., \citealt{chauvin}, \citealt{konopacky} \citealt{OFTI}, \citealt{logan}, \citealt{betapic}). 

The eccentricities of these long-period planets as an ensemble can provide important information about the dominant formation pathways for young giant planets. A natural methodology to determine the population-level eccentricity distribution underlying these samples is Hierarchical Bayesian Modeling (HBM), which allows joint modeling of the individual orbits and the population-level distribution \citep{Hogg}.  In particular, \citet{bowler:2020} (hereinafter BBN20) used this approach to analyze a sample of 27 imaged substellar companions (consisting of 9 giant planets and 18 brown dwarfs), and found that the giant planets in the sample had preferentially low eccentricities, while the brown dwarfs exhibited a broad range of eccentricities---implying that the dominant formation pathways for brown dwarf companions and giant planets between 5-100 AU are different. However, the functional forms of the recovered eccentricity distributions were sensitive to the choice of mass ratio or companion mass as a threshold to define the sample, hinting at the presence of small number statistical effects. The authors noted that while their results were sufficient to show that brown dwarfs and giant planets have qualitatively different underlying eccentricity distributions, forward-modeling experiments indicated that the data were not sufficient to constrain the exact shape of the population-level distributions.

In this work, we evaluate the reliability of using HBM to infer population-level eccentricity distributions of directly imaged planets, focusing in particular on the effect of hyperpriors on our ability to recover accurate underlying parameters. For this study, we adopt the Beta distribution to model the ensemble behavior of individual systems. 

The Beta distribution is a continuous probability distribution defined on [0,1] and has two shape parameters, $\alpha$ and $\beta$. This model has been frequently adopted for the purpose of inferring population-level eccentricity distributions (e.g., \citealt{kipping}, \citealt{shabram}, \citealt{beta_smallplanets}, BBN20, \citealt{tess_beta}). Most of these studies imposed uniform hyperpriors on the Beta distribution hyperparameters; to our knowledge, there has yet to be a systematic exploration of the impact that hyperprior choice has on HBM using the Beta distribution, either within or outside of the astronomical literature. Though previous studies have made use of alternate model choices to model eccentricity distributions such as the Rayleigh distribution and mixture models for eccentricity distribution with HBM \citep{beta_smallplanets}, we focus only on the Beta distribution in this study because we expect the small sample size of imaged planets will make it challenging to extract additional meaningful information that otherwise might be possible with a more detailed model comparison using a large sample.

This study is structured as follows. In Section \ref{epop}, we outline the HBM framewok we adopt for this analysis. We also present \texttt{ePop!}, a Python package for performing hierarchical modeling of eccentricities based on a sample of individual eccentricity distributions. Subsequent sections are structured as a series of experiments to assess the fidelity of various hyperpriors on both synthetic and real observations of imaged planets. In Section \ref{beta_hyperprior}, we use the behavior of the Beta Distribution to provide intuition and context for these effects. In Section \ref{Gaussian}, we isolate the impact of different hyperpriors by examining an idealised case for which the individual eccentricity posteriors are purely Gaussian. In Section \ref{forward_modelling_section}, we conduct a realistic forward modeling experiment simulating the observation and analysis process from astrometric measurements of individual systems to the reconstruction of an underlying eccentricity distribution. The aim of this exercise is to evaluate biases from the choice of hyperprior and the approach to orbit fitting using only small orbit arcs. Finally, in Section \ref{re-analysis}, we re-analyse the observations presented in BBN20 to test the impact of alternative hyperpriors on the inferred population-level eccentricity distribution of the widely separated giant planets uncovered by current direct imaging surveys. 

\section{\texttt{ePop!}} \label{epop}

To address the dearth of HBM software available to the astronomical community, we developed  \texttt{ePop!}\footnote{\texttt{ePop!} is available at \url{https://github.com/vighnesh-nagpal/ePop} under a 3-Clause BSD License, and Version 1.0 is archived in Zenodo \citep{ePop_code}.}, an open-source package written in Python for fitting population-level eccentricity distributions to sets of individual system eccentricity distributions. 

Hierarchical modeling is widely used to simultaneously determine Bayesian posteriors for individual objects in a sample as well as posteriors over the population-level parameters for an assumed model. This approach is inefficient for our science case, where calculating individual system orbital posteriors for directly imaged companions can be time-consuming and computationally non-trivial when only short orbit arcs are available (e.g. \citealt{OFTI}). Moreover, for directly imaged planets and brown dwarfs, orbit posteriors are in some cases already available and do not need to be recomputed. 

\citet{Hogg} developed an approximation to the HBM likelihood that makes use of precomputed samples from individual system posteriors. This procedure separates the steps of inferring individual and population-level posteriors. We use this approach to define our hierarchical likelihood as:
\begin{equation}
    \label{hierarchicalmodel}
    \mathcal{L}_{\textbf{v}} \approx \prod_{i=1}^{N} \frac{1}{K} \sum_{j=1}^{K_{i}} B_{\textbf{v}}(e_{ij}),
\end{equation}
where \textit{N} is the number of systems under consideration, $K_{i}$ is the number of samples contained within the $i^{}$th individual eccentricity posterior, $e_{ij}$ is the $j^{}$th eccentricity sample for the $i^{}$th system, and \textbf{v} is the vector of population-level model parameters, defined in our nominal model to be: 
\begin{gather}
    \textbf{v}=(\alpha,\beta),
\end{gather}
where $\alpha$, $\beta$ are the Beta distribution hyperparameters. We assume that the eccentricity prior on each individual system is uniform. The Beta distribution, $B$, is defined as: 
\begin{equation}
    \label{beta_equation}
      B_{\textbf{v}}(e)=\frac{\Gamma(\alpha+\beta)}{\Gamma(\alpha) \Gamma(\beta)}e^{\alpha-1}(1-e)^{\beta-1},  
\end{equation} where $\Gamma$ is the Gamma function. Using the affine-invariant Markov Chain Monte Carlo (MCMC) sampler {\fontfamily{pcr}\selectfont
emcee} \citep{emcee}, \texttt{ePop!} computes the posterior of the hyperparameter vector \textbf{v} following the likelihood function defined in Equation \ref{hierarchicalmodel}. In addition, \texttt{ePop!} contains functionality for applying  different hyperpriors on the hyperparameters $\alpha$ and $\beta$, in the form of \textbf{Prior} objects in the code. Examples of families of hyperpriors that can be used are summarised in Table \ref{table:epop_priors}. In the code block below, \texttt{ePop!} is used to compute the posterior of the hyperparameters for a set of individual eccentricity posteriors using a log-uniform hyperprior.

\begin{lstlisting}[language=Python]
import ePop 
import glob
    
fnames=sorted(glob.glob('./posteriors/*'))

# load individual eccentricity distributions
posts=[np.load(f) for f in fnames]

# create Likelihood object and choose prior
like=ePop.hier_sim.Pop_Likelihood(posteriors=posts,prior='log_uniform')

# sample the hyperparameters using MCMC
beta_samples=like.sample(2000,burn_steps=500,nwalkers=30)

\end{lstlisting}

\texttt{ePop!} currently uses the Beta distribution for hierarchical modeling, but it can be expanded to include other parametric model families in the future.


\begin{deluxetable*}{ccc}
\tabletypesize{\footnotesize}
\tablecolumns{3}
\tablewidth{50pt}
\tablecaption{Hyperpriors in {\fontfamily{pcr}\selectfont
e-Pop!} \label{table:epop_priors}}
\tablehead{
 \colhead{Prior} & \colhead{Parameters} & \colhead{Functional Form}
}
\startdata
Uniform & Lower Bound: \textit{x$_{0}$}, Upper Bound: \textit{x$_{1}$} & 
p($x$) = $\begin{cases} \frac{1}{x_{1}-x_{0}},& \text{if } x\in [x_0,x_1]\\ 0 & 0,               \text{otherwise} \end{cases}$ \\
Truncated Gaussian & Mean: $\mu$, Standard Deviation: $\sigma$ & 
p($x$) = $\begin{cases} \frac{1}{\sigma}\frac{\frac{1}{\sqrt{2\pi}} e^{-\frac{1}{2}\left(\frac{x-\mu}{\sigma}\right)^{2}}}{1-\frac{1}{2} \left(1+\mathrm{erf} \left(\frac{-\mu}{\sigma \sqrt{2}}\right)\right)},& \text{if } x > 0\\ 0 & \text{otherwise} \end{cases}$ \\
Log-uniform & Lower Bound: \textit{x$_{0}$}, Upper Bound: \textit{x$_{1}$} &
p($x$) = $\begin{cases} \frac{1}{x}\ ,& \text{if } x \in [x_0, x_1]\\ 0 & \text{otherwise} \end{cases}$
\\
Log-normal & Log-space Mean:$\mu$, Log-space Standard Deviation $\sigma$ & 
p($x$) = $\begin{cases} \frac{1}{x \sigma \sqrt{2 \pi}} e^{ -\frac{\mathrm{ln}(x)-\mu ^2}{2{\sigma}^2}},& \text{if } x > 0\\ 0, & \text{otherwise} \end{cases}$
\\
Gamma & Shape: $\alpha$, Rate: $\beta$ &
p($x$) = $\begin{cases} \frac{{\beta}^{\alpha}}{\Gamma ({\alpha})} x^{\alpha-1} e^{-\beta x} ,& \text{if } x > 0\\ 0, & \text{otherwise} \end{cases}$
\\
\enddata
\end{deluxetable*}

\section{On hyperpriors for Beta distributions} \label{beta_hyperprior}

Many studies using the Beta Distribution for hierarchical Bayesian modeling have imposed uniform priors on the hyperparameters $\alpha$ and $\beta$ (e.g. \citealt{beta_smallplanets}, BBN20, \citealt{tess_beta}). However, few studies have investigated whether uniform hyperpriors on the Beta distribution's shape parameters actually correspond to an \textit{uninformative} (or reasonably weakly informative) prior on the underlying eccentricity distribution. For the context of this study, we seek to identify hyperpriors that result in a family of broad and flexible eccentricity distributions, with a preference for wider, more uniform distributions. Physically, this preference is motivated by the desire to encompass a variety of different planetary formation models. As an example, we would like to equally weight zero-peaked distributions characteristic of formation in a disk \citep{armitagebook} and the broader distributions characteristic of outward dynamical ejections \citep{veras}.

\begin{figure*}
    \centering
    \includegraphics[width=\textwidth]{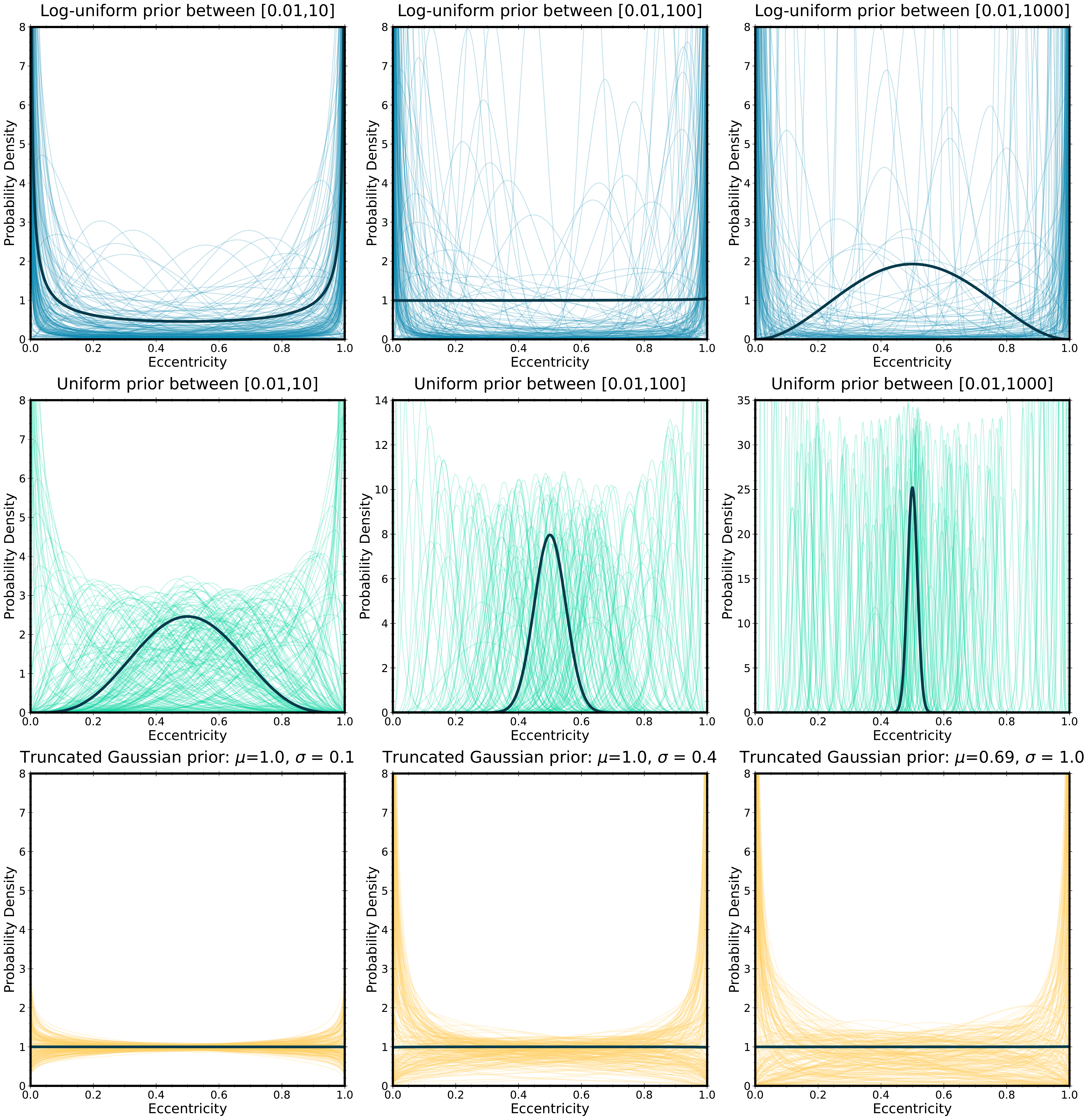}
    \caption{Examples of families of hyperprior eccentricity distributions randomly drawn from different hyperpriors of the Beta distribution shape parameters $\alpha$ and $\beta$. Each panel shows a representative sample of 200 eccentricity distributions randomly drawn from a given hyperprior. The eccentricity distribution corresponding to the median ($\alpha$, $\beta$) drawn from the hyperprior is highlighted in bold. The top row shows the distributions drawn from a log-uniform hyperprior with lower bound of 0.01 and upper bounds of 10, 100 and 1000 (from left to right). The middle row shows distributions drawn from uniform hyperpriors spanning ranges of [0.01,10], [0.01,100], and [0.01,1000]. These have a tendency to produce progressively narrower distributions as the upper bound increases. In the third row, we show the distributions drawn from three different truncated Gaussian hyperpriors. From left to right, these hyperpriors are parameterized by: ($\mu, \sigma$) = (1.0, 0.1), ($\mu, \sigma$) = (1.0, 0.4), and ($\mu, \sigma$) = (0.69, 1.0).}
    \label{fig:visualisation}
\end{figure*}

To visualize what uniform, log-uniform, and (truncated) Gaussian\footnote{For the truncated Gaussian case, we draw from a normal distribution but discard samples where $\alpha$ or $\beta$ are negative, for which the Beta Distribution is undefined. See Table \ref{table:epop_priors} for the equation for the truncated Gaussian functional form.} hyperpriors on the Beta distribution parameters correspond to in eccentricity space, we draw 1000 samples of ($\alpha$, $\beta$) from each hyperprior distribution (assuming no covariance between $\alpha$ and $\beta$) and then plot the corresponding eccentricity distributions in Figure \ref{fig:visualisation}. Nine permutations of hyperpriors are examined in total: three each for the uniform, log-uniform and truncated Gaussian functional forms.  In the case of the uniform and log-uniform hyperpriors, we test three sets of bounds on ($\alpha$, $\beta$): [0.01, 10], [0.01, 100], and [0.01, 1000]. For the case of the truncated Gaussian functional form, we examine the impact that hyperprior width has on the range of eccentricity distributions it can produce by testing three values of $\sigma$: 0.1, 0.4, and 1.0. For each $\sigma$, we choose the value of $\mu$ for which the median of the hyperprior parameterized by ($\mu$,  $\sigma$) is 1, a decision motivated by the fact that $\alpha=\beta=1$ corresponds to a uniform eccentricity distribution in eccentricity space. Following this process yields truncated Gaussian hyperpriors parameterized by ($\mu, \sigma$) = (1.0, 0.1), (1.0, 0.4), and (0.69, 1.0). Further details regarding the hyperpriors we test are summarized in Table \ref{table:tested_hyperpriors}.

As seen in Figure \ref{fig:visualisation}, sampling from a uniform hyperprior on the Beta distribution parameters with bounds [0.01,1000] returns distributions that are generally narrowly peaked. This behavior can be explained by the nature of how $\alpha$ and $\beta$ impact the Beta distribution:

\begin{itemize}
    \item $\alpha,\beta > 1$: Higher values of the Beta parameters correspond to more narrowly peaked distributions. The ratio $\frac{\alpha}{\beta}$ determines the location of the peak.
    
    \item $\alpha< 1$: Asymptotic at 0.
    
    \item $\beta< 1$: Asymptotic at 1.
    
    \item $\alpha=\beta=1$: Uniform, flat distribution.
    
\end{itemize}  

\begin{deluxetable}{ccc}
\tabletypesize{\footnotesize}
\tablecolumns{4}
\tablewidth{\textwidth}
\tablecaption{Beta distribution parameter values for nine different hyperpriors. See Figure \ref{fig:visualisation} for a visualization of their corresponding eccentricity distributions. {\fontfamily{pcr}\selectfont
} \label{table:tested_hyperpriors}}
\tablehead{
 \colhead{Hyperprior} & \colhead{Parameter Median} & \colhead{$68\%$ Interval}
}
\startdata
Uniform on [0.01,10] &   $5.0$ & $[1.6,6.4]$\\
Uniform on [0.01,100] &   $50$ & $[16,64]$ \\
Uniform on [0.01,1000] &  $500$ & $[160,640]$ \\ 
Log-uniform on [0.01,10] & $0.32$ & $[0.03,3.32]$ \\
Log-uniform on [0.01,100] & $1.00$ & $[0.05,30.89]$ \\
Log-uniform on [0.01,1000] & $3.16$ & $[0.06,158.62]$ \\
Truncated Gaussian: $\mu=1.0$, $\sigma=0.1$ & 
$1.00
$ & $[0.91,1.09]$ \\
Truncated Gaussian: $\mu=1.0$, $\sigma=0.4$ & 
$1.00$ & $[0.61,1.40]$ \\
Truncated Gaussian: $\mu=0.69$, $\sigma=1.0$ & $1.00$ & $[0.45,2.12]$ \\
\enddata
\end{deluxetable}

In cases where the observational sample size is small or individual orbits are poorly constrained (such as in the current sample of imaged planets), the linearly uniform hyperprior distributions are biased towards narrow eccentricity distributions, and can impart this feature onto the derived population-level distributions. Moreover, for a uniform hyperprior distribution in $\alpha$ and $\beta$, the broader the adopted range for each parameter, the more narrowly constrained the family of hyperprior distributions will be in eccentricity space.  This can be seen by looking at the bounds; a range of (0,10] in $\alpha$ and $\beta$ (with mean values of 5 in each hyperparameter) will return a broader set of distributions than a range of (0,100] (with mean values of 50).  This is opposite to the usual, more intuitive sense of uniform priors on unknown parameters in Bayesian inference: in this case the broader the prior range on hyperparameters, the more restricted the resulting hyperprior distributions will be in physical space.

On the other hand, sampling from log-uniform hyperpriors between [0.01,100] or the specific truncated Gaussian hyperpriors listed in Table \ref{beta_hyperprior} returns a family of distributions for which the median is a uniform eccentricity distribution. We wish to find hyperpriors which do not impart a systematic bias towards any specific shape and are also capable of producing a variety of flexible behaviors; among those we examine, both the Gaussian hyperprior with $\mu$=0.69, $\sigma=1.0$ and log-uniform hyperprior between [0.01,100] encompass eccentricity distributions with a wide variety of qualitative shapes (in the process capturing many of the morphologies predicted by different models of planet formation and orbit evolution) and have median distributions that are close to uniform in eccentricity---qualities that stand out as promising candidates for use as hyperpriors in HBM. 

\section{Experiment with Gaussian Posteriors} \label{Gaussian}

\subsection{Method}

To explore the impact that the choice of hyperpriors has on the population parameter posteriors, we follow the approach of BBN20 (see their Section 4.3.1) in which synthetic individual posteriors are drawn from a known underlying eccentricity distribution. For this experiment, we randomly draw $N$ eccentricity values from an underlying `true' population-level distribution and, for the purposes of this exercise, assume Gaussian eccentricity posteriors centered on each draw with standard deviation $\sigma$. Here, \textit{$N$} represents the number of objects in an observational sample and $\sigma$ describes how well each object's orbit is constrained. 

Once the set of posteriors is constructed, we carry out the hierarchical modeling with \texttt{ePop!} to sample the posterior parameters over population-level hyperparameters. We conduct three experiments, each imposing a different hyperprior on ($\alpha$, $\beta$). In the first, we test a uniform hyperprior on the interval $[0.01,1000]$, mirroring the choices made by BBN20 in order to facilitate  comparison with their results. For the second fit, we impose a truncated Gaussian hyperprior with $\mu=0.69, \sigma=1.0$, and for the third, we impose a log-uniform prior with bounds $[0.01,100]$. These specific shape parameters and ranges were selected so that the median values of $\alpha$ and $\beta$ are both 1.0 for experiments 2 and 3, which correspond to a flat distribution in eccentricity space. To quantify how well a given fit recovers the true underlying eccentricity distribution, we calculate a normalized residual metric $\mathcal{M}$:

\begin{equation}
    \label{metric}
    \mathcal{M} = \frac{1}{J} \sum_{i=1}^{J}
    \int_{0}^{1} {|B_{\textbf{v}_i}(e)-B_{\textbf{v}_u}(e)|} de, 
\end{equation}

\noindent where $J$ is the number of eccentricity distributions randomly drawn from the Beta distribution hyperparameter posteriors. $\mathcal{M}$ is the average area of the absolute deviation between the `true' distribution $(B_{\textbf{v}_u})$ and the family of posterior eccentricity distributions. Smaller values of $\mathcal{M}$ indicate fits that better recover the underlying eccentricity distribution: $\mathcal{M}=0$ corresponds to the limiting case in which the underlying distribution is perfectly recovered. In the context of this study, we use $\mathcal{M}$ to assess the relative performance of different hyperpriors with respect to how accurately they recover known underlying eccentricity distributions.

\begin{figure*}
    \centering
    \includegraphics[width=\textwidth]{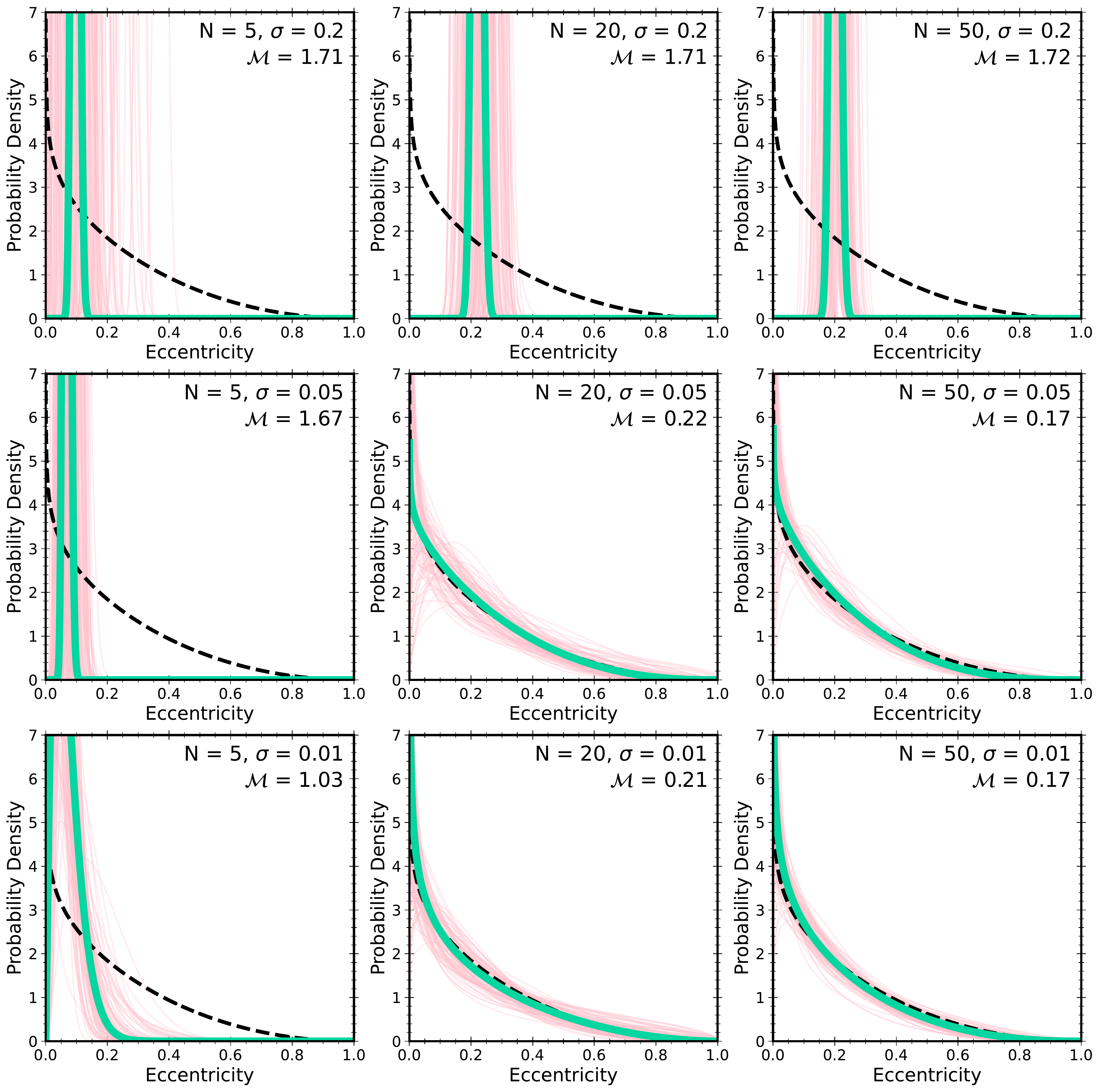}
    \caption{Results of hierarchical Bayesian fitting using \textit{uniform hyperpriors} for $\alpha$ and $\beta$ on [0.01,100] for a set of synthetic Gaussian eccentricity posteriors drawn from the RV exoplanet distribution (dashed line). The 9 panels show the posteriors of population-level distributions obtained as \textit{N} (sample size or number of artificial planets in the analysed dataset) and $\sigma$ (width of the individual eccentricity posteriors) are varied. The thick solid curves are the median recovered distributions while the thin lines are randomly sampled distributions from the posterior. In the case of low $N$ or high $\sigma$, the recovered distributions are preferentially narrowly peaked, a trend reminiscent of the results found for the giant planet sample in BBN20. Furthermore, we see that this trend persists across the first row, even when N increases by a factor of 10, a sign that the results are being biased by the choice of a uniform prior. }
    \label{fig:uniform}
\end{figure*}

\begin{figure*}
    \centering
    \includegraphics[width=\textwidth]{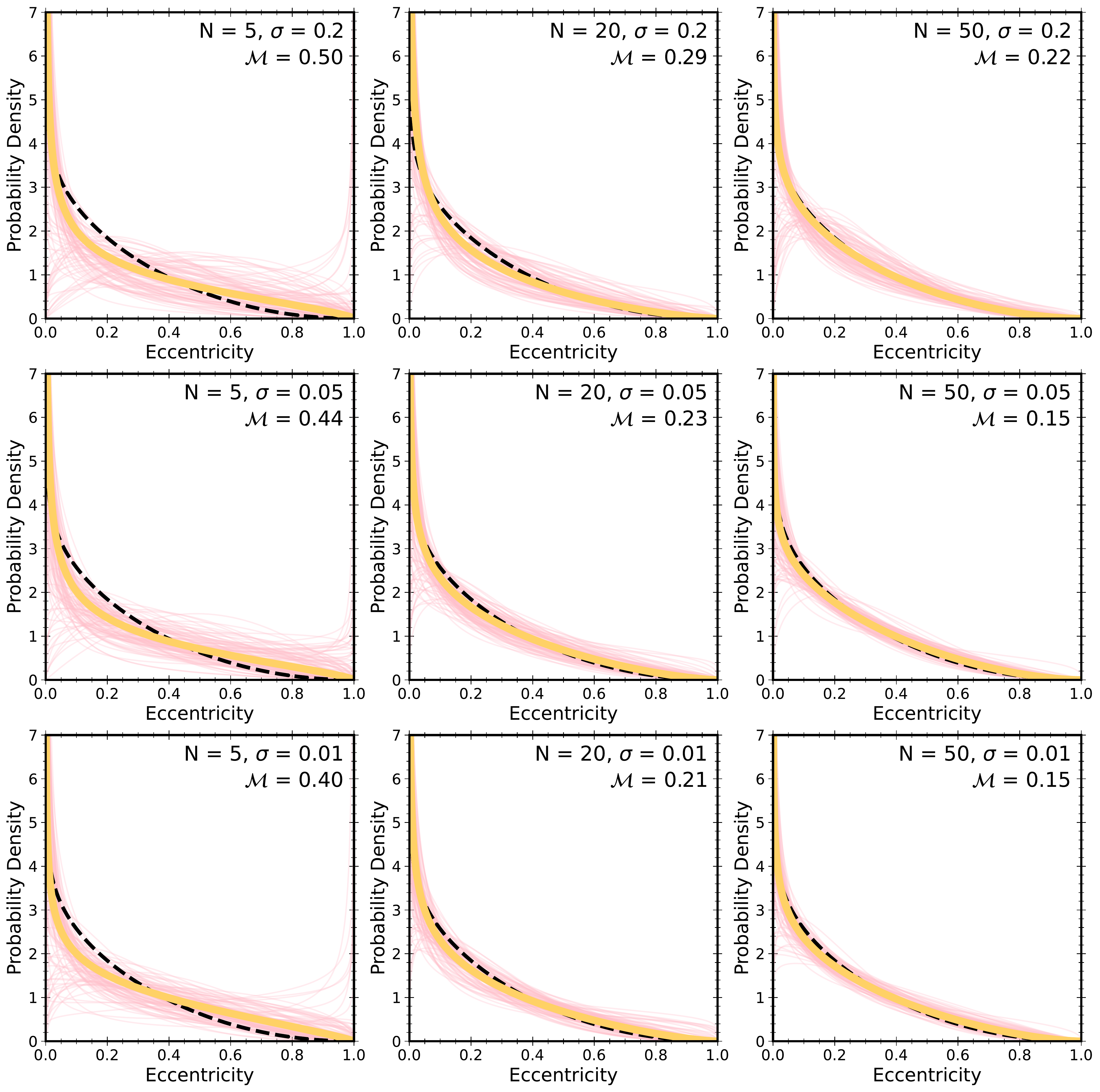}
    \caption{Analogous to Figure \ref{fig:uniform}, but here we impose \textit{a truncated Gaussian hyperprior} with $\mu=0.69$ and $\sigma=1.0$ on ($\alpha$, $\beta$). In contrast with the case of the uniform hyperprior seen in Figure \ref{fig:uniform}, even in the case of low $N$ or high $\sigma$, the distributions recovered are reminiscent of the underlying RV exoplanet eccentricity distribution. The preference for narrowly peaked distributions observed when imposing a uniform hyperprior disappears in this case, and the posteriors of inferred distributions approach the underlying distribution as $N$ increases and $\sigma$ decreases.}
    \label{fig:gaussian}
\end{figure*}

\begin{figure*}
    \centering
    \includegraphics[width=\textwidth]{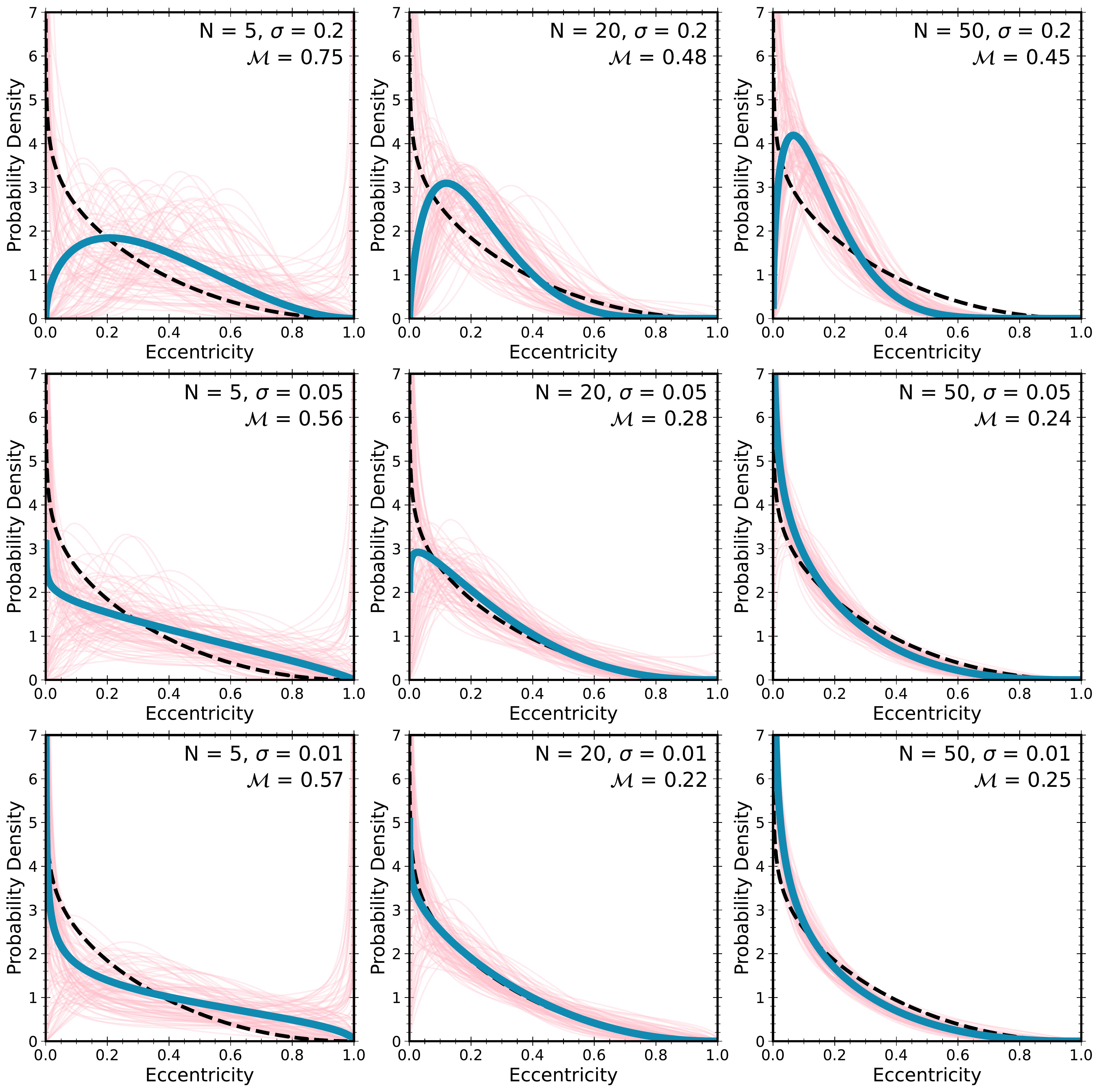}
    \caption{Analogous to Figure \ref{fig:uniform}, but here we impose a \textit{log-uniform hyperprior} with bounds [0.01,100] on ($\alpha$,$\beta$). Similar to what we observe in Figure \ref{fig:gaussian} for truncated Gaussian hyperpriors, there is no artificial preference for narrowly peaked distributions. This provides further evidence that the narrow peaks for small samples in Figure \ref{fig:uniform} for uniform hyperpriors arise from prior-driven effects.}
    \label{fig:loguniform}
\end{figure*}

\subsection{Results}

The results of the fits using the uniform, truncated Gaussian, and log-uniform hyperpriors are shown in Figures \ref{fig:uniform}, \ref{fig:gaussian},  and \ref{fig:loguniform}, respectively.  For these tests, we adopt the radial velocity (RV) exoplanet eccentricity distribution from \citet{kipping}, for which $\alpha=0.87$ and $\beta=3.03$. The nine panels in each figure display how the results change when the number of systems ($N$) is varied from $5,20,$ and $50$, and $\sigma$ is varied from $0.2,0.05$, and $0.01$. For each panel, we compute and include the corresponding $\mathcal{M}$ metric value. As seen in Figure \ref{fig:uniform}, using a uniform hyperprior reliably recovers the input distribution in the best case scenario ($N=50$, $\sigma=0.01$), but fails when $N$ is small (first row) or when the individual eccentricity posteriors are less constraining (high $\sigma$, first column).

In particular, the recovered distributions in these cases tend to be narrowly peaked and show little improvement across the first row or the first column, even when $N$ increases to $50$ or $\sigma$ improves to $0.01$. Similar behavior was observed by BBN20 in their version of this experiment. The small size of the current sample of imaged giant planets means that attempts to infer its underlying eccentricity distribution may be impacted by the choice of hyperprior.

On the other hand, using the truncated Gaussian hyperprior with $\mu=0.69,  \sigma=1.0$ results in marked improvement in our ability to recover the underlying distribution, even when contending with broad individual posteriors and small sample sizes, as seen in Figure \ref{fig:gaussian}. We observe a similar level of improvement when we impose the log-uniform hyperprior with bounds [0.01,100]. The recovered distributions in the first row and column of both Figures \ref{fig:gaussian} and \ref{fig:loguniform} do not exhibit the same narrow peaks seen in the case of uniform hyperpriors. Even in the the worst case from our experiment ($N=5$, $\sigma=0.2$), both hyperpriors manage to recover distributions qualitatively consistent with the input RV exoplanet distribution.

\section{Forward Modeling} \label{forward_modelling_section}

\subsection{Method} \label{forward_modelling_method}
\begin{figure}
    \centering
    \includegraphics[width=0.5\textwidth]{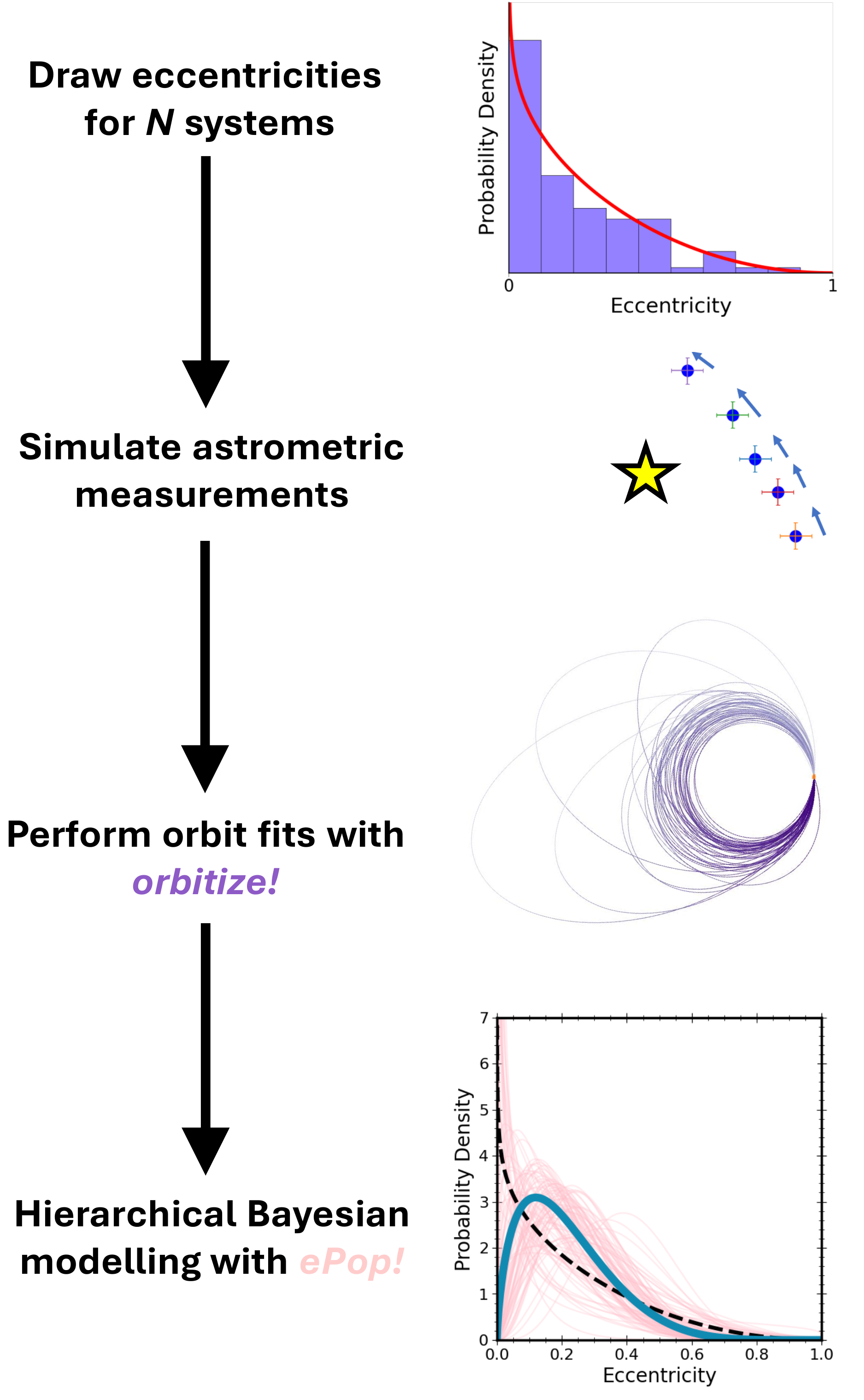}
    \caption{Flowchart summarizing the forward modeling process described in Section \ref{forward_modelling_method}. We begin by assuming an underlying eccentricity distribution, from which we randomly sample $N$ systems. Then for each system, we use a Keplerian model to simulate 5 evenly spaced astrometric points. Orbit fits are performed for each system using \texttt{OFTI} which produce a set of $N$ eccentricity posteriors. Finally, we use \texttt{ePop!} to run a hierarchical fit on this set of posteriors, producing a posterior on the hyperparameters $\alpha$ and $\beta$, which translates to a posterior of underlying population-level eccentricity distributions.}
    \label{fig:flowchart}

\end{figure}

\begin{figure}
    \centering
    \includegraphics[width=0.44\textwidth]{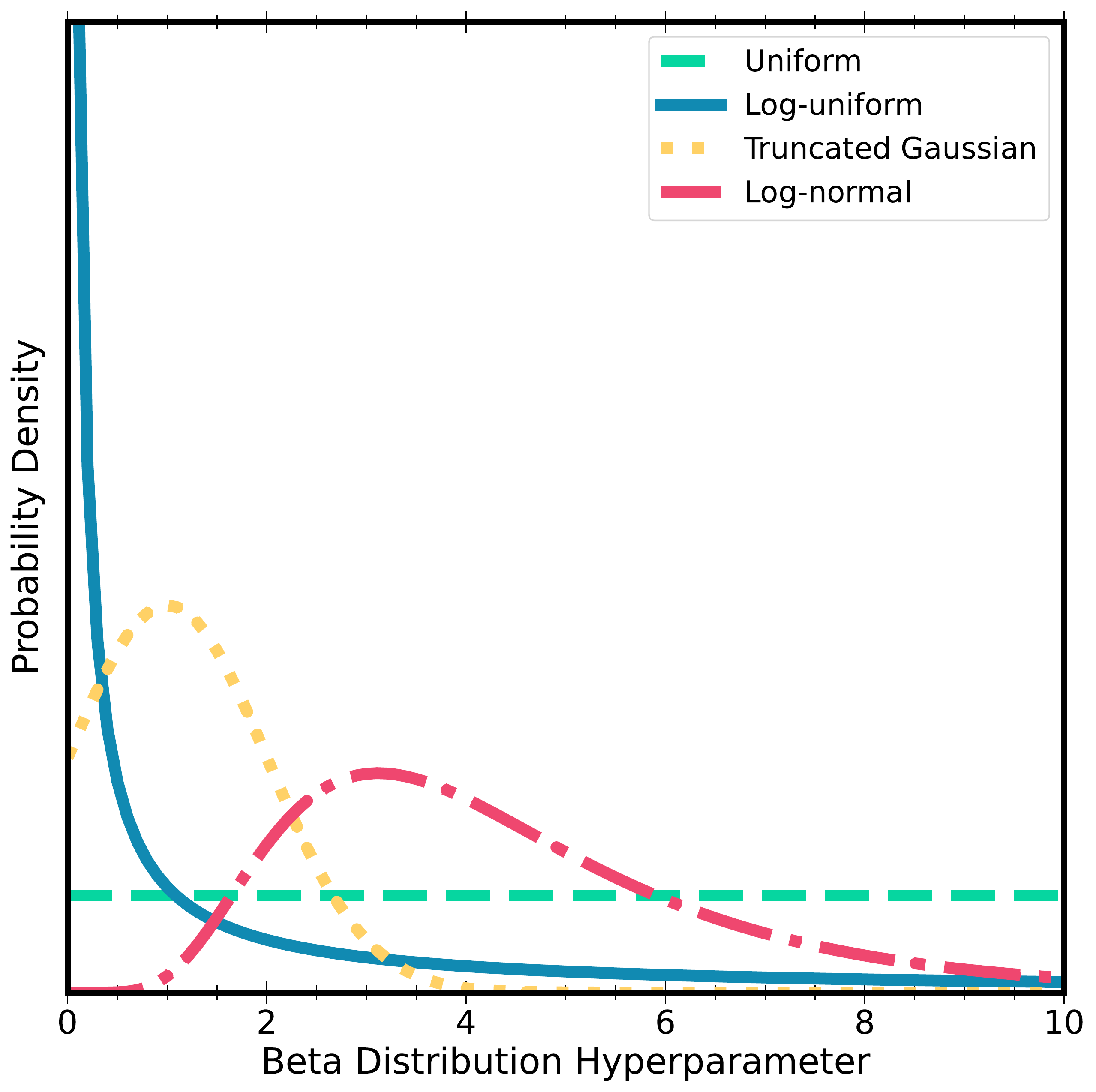}
    \caption{The hyperpriors imposed on both $\alpha$ and $\beta$ for the hierarchical fitting of our forward modeled dataset. We chose these distributions to test the hyperprior impact on the recovered distribution. We only plot hyperprior distributions for hyperparameter values $\leq 10$ for ease of displaying the qualitative distinctions between the hyperpriors.}
    \label{fig:hyperprior}

\end{figure}

To test whether the findings of Section \ref{Gaussian} persist under realistic conditions, we developed a method for testing the overall validity of hierarchical eccentricity modeling for the directly imaged population of substellar objects by carrying out a series of end-to-end tests. We simulated the process of taking data, fitting orbits, and recovering the underlying distribution to assess how similar the recovered distribution was to the underlying distribution. An illustration of this process is shown in Figure \ref{fig:flowchart}. Specifically, this entailed:

\begin{itemize}
    \item Generating a sample: We begin by assuming an underlying population-level distribution (either uniform or the RV exoplanet distribution from \citealt{kipping}) from which we then sample $N$ eccentricities (corresponding to individual systems). We assume that each system has a host star of 1 \(M_\odot\) and draw semi-major axes from a log-uniform distribution between 10-100 AU, which is within the 1-sigma results of \citet{nielsen}. We draw inclinations from a  distribution uniform in $\cos{i}$ to account for the isotropic distribution of exoplanet orbital inclinations. 
    
    \item Simulating astrometry and fitting orbits: For each system, we use a Keplerian model to simulate five astrometric points evenly spaced over an observational window of 2000 days. To mimic observational uncertainties, we add Gaussian noise to the simulated astrometry, and then use the Orbits for the Impatient (OFTI, \citealt{OFTI})  implementation in \texttt{orbitize!} \citep{orbitize} to sample the orbital posterior distribution.

    \item Hierarchical Bayesian Modeling: We then use \texttt{ePop!} to sample the posterior over population-level parameters of the Beta distribution model, given the set of $N$ eccentricity posteriors from the orbit fits. For each set of posteriors, we perform this fit four times, applying morphologically different hyperpriors on ($\alpha$, $\beta$) in each case to test their impact on the final results. For this purpose, we use the following hyperpriors : truncated Gaussian with $\mu=0.69$, $\sigma=1.0$, uniform on [0.01,100], log-uniform on [0.01,100], and log-normal with $\mu=4.0$, $\sigma=0.5$. See Figure \ref{fig:hyperprior} for a visualization of these hyperpriors. We chose the log-normal hyperprior (which has a non-uniform mean eccentricity distribution) to investigate the effects
    of imposing a strongly biased hyperprior on the recovered distributions.

\end{itemize}

\begin{figure*}
    \centering
    \includegraphics[width=\textwidth]{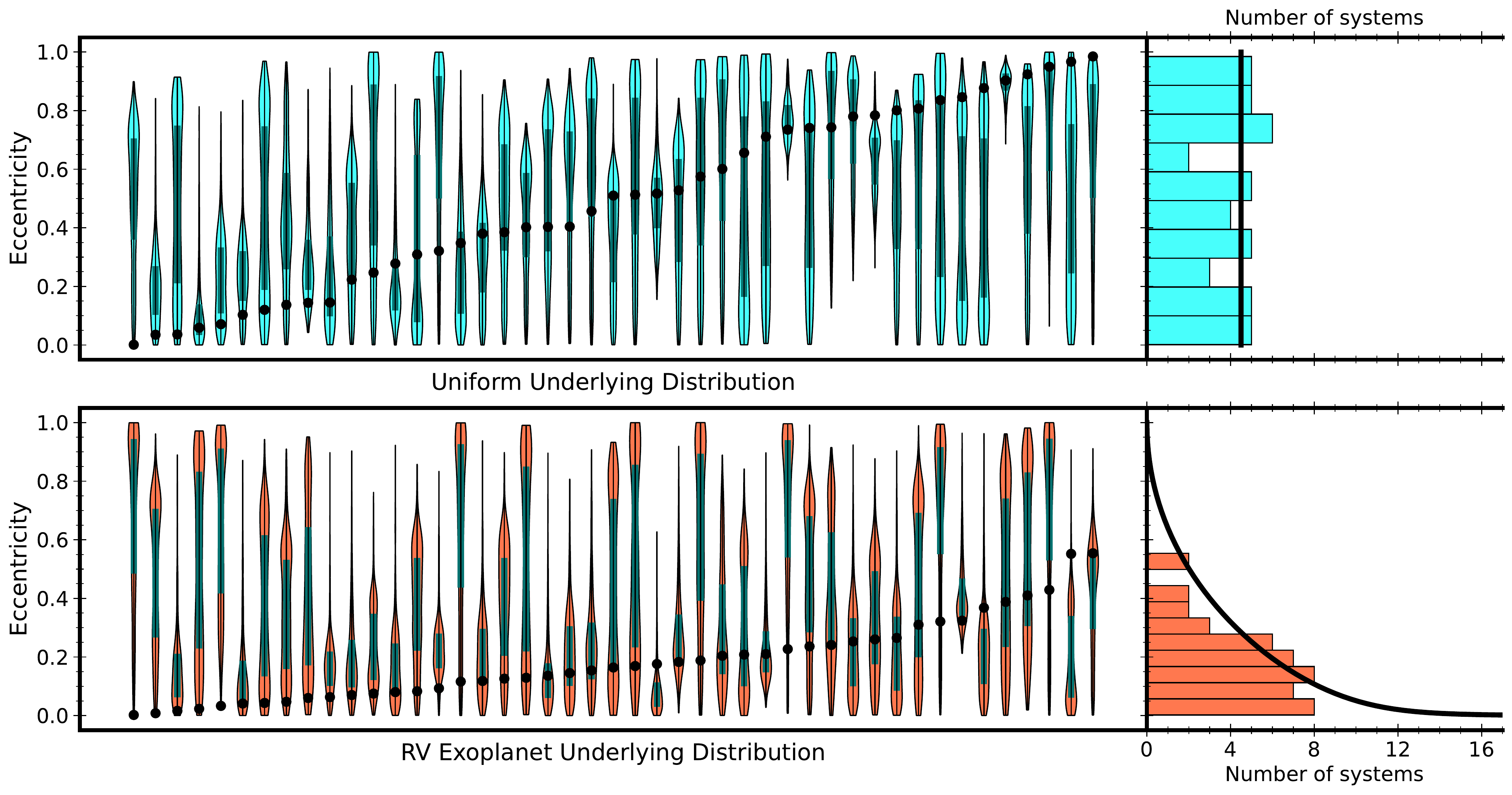}
    \caption{Visualization of the simulated forward modeled sample. In the top panel, we show the individual eccentricity posteriors (violin plots) obtained by using \texttt{OFTI} to perform orbit fits to the simulated astrometry for each of the 45 systems drawn from an underlying uniform population-level eccentricity distribution. In each individual violin plot, the black dot corresponds to the 'true' eccentricity of the individual system, while the gray box corresponds to the interquartile range of its eccentricity posterior. On the right, we plot the histogram of true eccentricities for the simulated sample, where the thick black curve shows the underlying uniform eccentricity distribution. The bottom panel shows the analogous information for the ensemble of 45 systems drawn from an underlying RV exoplanet distribution. }
    \label{fig:tower}

\end{figure*}

\subsection{Results} \label{cosi}

We focus our analysis on two choices of underlying models: a uniform eccentricity distribution, and the RV exoplanet distribution---a choice motivated by the open question of whether it is possible to recover distributions that we may expect a-priori, given the complications introduced by incorporating realistic observing conditions and uncertainties. For each case, we first simulate a set of systems following the procedure outlined in Section \ref{forward_modelling_method}, and then perform orbit fits on sub-samples of size $N$=5, 10, 20,  and 45. Representative individual eccentricity posteriors (with true eccentricities highlighted) from these fits for both the uniform and RV exoplanet eccentricity distributions are shown in Figure \ref{fig:tower}. We then perform hierarchical fits to these sub-samples using the four different hyperpriors previously discussed, which are shown in Figure \ref{fig:hyperprior}. 

\begin{figure*}
    \centering
    \includegraphics[width=\textwidth]{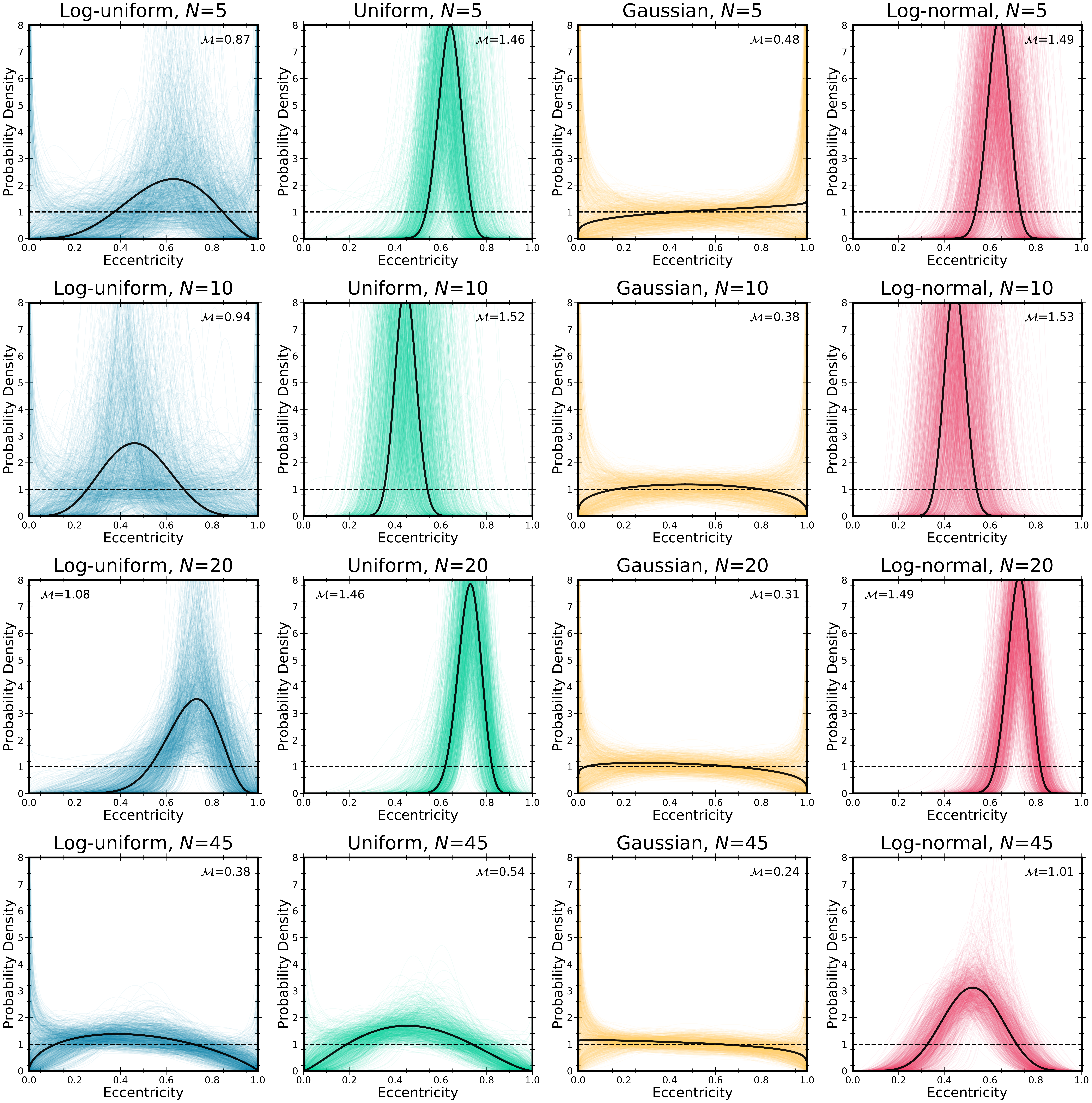}
    \caption{Results of applying the forward modeling process described in Section \ref{forward_modelling_method} to simulated samples of imaged companions drawn from an underlying uniform eccentricity distribution. Four sample sizes ($N$=5, 10, 20, and 45) and four hyperpriors (Log-uniform, Uniform, Gaussian, Log-normal) are tested. In each panel, we show randomly drawn distributions from the posterior on ($\alpha$, $\beta$) and highlight the eccentricity distribution corresponding to the median recovered hyperparameter values in bold. The corresponding value of $\mathcal{M}$ quantifies the level of agreement between the recovered and input underlying distributions. For reference, we plot the underlying eccentricity distribution (in this case, the uniform distribution) as a dashed line. For each sample size, the Gaussian hyperprior performs the best, followed by the log-uniform, uniform, and log-normal hyperpriors. The uniform, log-uniform and log-normal hyperpriors impose a strong bias towards narrowly peaked distributions that is most evident for small sample sizes.}
    \label{fig:uniform_cosinc}

\end{figure*}

For the case where we forward model an input uniform eccentricity distribution (Figure \ref{fig:uniform_cosinc}), we find that the Gaussian hyperprior consistently performs the best out of the hyperpriors tested; this hyperprior consistently produced the lowest $\mathcal{M}$ values for every sample size $N$. We further observe that the hyperpriors follow a clear hierarchy across sample sizes: the Gaussian performs the best, followed by the log-uniform, uniform, and then log-normal---a picture broadly in-line with the results of Section \ref{Gaussian}. See  Table \ref{table:forward_uniform} for a summary of the recovered posteriors on ($\alpha$, $\beta$) and corresponding values of $\mathcal{M}$ for each fit we perform as part of this test case.

Although convergence to the true distribution is reached fairly quickly for the truncated Gaussian case, there remains significant power at high and low eccentricities. This may be caused by biases that arise when fitting systems with low orbital coverage. For example, the inclination of a system can exert great influence on estimates of orbital eccentricity, an effect that can manifest as posteriors that are skewed away from the true eccentricity, biased towards high or low eccentricity, or bimodal (\citealt{rodrigo}). The compounding of such biases in individual eccentricity posteriors, combined with the effects of measurement uncertainty may be responsible for HBM's struggles with inferring the exact underlying population level distributions, even for large sample sizes.

\begin{deluxetable}{ccccc}
\tabletypesize{\footnotesize}
\tablecolumns{4}
\tablewidth{\textwidth}
\tablecaption{Values of the recovered hyperparameters ($\alpha$, $\beta$) and metric $\mathcal{M}$ for forward modeling a uniform underlying distribution ($\alpha=1$, $\beta=1$) following the procedure outlined in Section \ref{forward_modelling_method}. The quoted uncertainties correspond to the 68\% credible intervals for $\alpha$ and $\beta$. See Figure \ref{fig:uniform_cosinc}  for a visualisation of these distributions. For each sample size $N$, fits using the truncated Gaussian recover the most accurate family of population-level distributions, as evidenced by the small values of these fits' corresponding $\mathcal{M}$ metrics (in bold). {\fontfamily{pcr}\selectfont
} \label{table:forward_uniform}}
\tablehead{
 \colhead{Hyperprior} & \colhead{$N$} & \colhead{$\alpha$} & \colhead{$\beta$} & \colhead{$\mathcal{M}$}
}
\startdata
Uniform on [0.01,1000] & 5 &  $59${\raisebox{0.5ex}{\tiny$\substack{+29 \\ -33}$}} & $34${\raisebox{0.5ex}{\tiny$\substack{+21 \\ -19}$}} & 1.47 \\
Log-uniform on [0.01,100] & 5 & $4.7${\raisebox{0.5ex}{\tiny$\substack{+15 \\ -3.5}$}} & $3.2${\raisebox{0.5ex}{\tiny$\substack{+8.2 \\ -2.1}$}} & 0.85 \\
Truncated Gaussian: $\mu=0.69$, $\sigma=1.0$ & 5 & $1.24${\raisebox{0.5ex}{\tiny$\substack{+0.45 \\ -0.42}$}} & $0.56${\raisebox{0.5ex}{\tiny$\substack{+0.46 \\ -0.25}$}} &\textbf{ 0.61} \\
Log-normal: $\mu=1.0$, $\sigma=0.4$ & 5  & $63${\raisebox{0.5ex}{\tiny$\substack{+26 \\ -31}$}} & $36${\raisebox{0.5ex}{\tiny$\substack{+20 \\ -18}$}} & 1.49 \\
Uniform on [0.01,1000] & 10 &
$53${\raisebox{0.5ex}{\tiny$\substack{+25 \\ -24}$}} & $66${\raisebox{0.5ex}{\tiny$\substack{+24 \\ -30}$}} & 1.52 \\
Log-uniform on [0.01,100] & 10 &
$5.7${\raisebox{0.5ex}{\tiny$\substack{+24 \\ -4.3}$}} & $6.5${\raisebox{0.5ex}{\tiny$\substack{+31 \\ -5.1}$}} & 0.91   \\
Truncated Gaussian: $\mu=0.69$, $\sigma=1.0$ & 10 &
$1.1${\raisebox{0.5ex}{\tiny$\substack{+0.30 \\ -0.34}$}} & $1.1${\raisebox{0.5ex}{\tiny$\substack{+0.40 \\ -0.35}$}} & \textbf{0.28 }\\
Log-normal: $\mu=1.0$, $\sigma=0.4$ & 10 &
$53${\raisebox{0.5ex}{\tiny$\substack{+25 \\ -23}$}} & $66${\raisebox{0.5ex}{\tiny$\substack{+23 \\ -28}$}} & 1.54 \\
Uniform on [0.01,1000] & 20 & 
$56${\raisebox{0.5ex}{\tiny$\substack{+30 \\ -32}$}} & $21${\raisebox{0.5ex}{\tiny$\substack{+12 \\ -12}$}} & 1.44  \\
Log-uniform on [0.01,100] & 20 &
$11${\raisebox{0.5ex}{\tiny$\substack{+20 \\ -6.8}$}} & $4.8${\raisebox{0.5ex}{\tiny$\substack{+7.7 \\ -2.6}$}} & 1.06 \\
Truncated Gaussian: $\mu=0.69$, $\sigma=1.0$ & 20 &
$1.00${\raisebox{0.5ex}{\tiny$\substack{+0.37 \\ -0.31}$}} & $1.18${\raisebox{0.5ex}{\tiny$\substack{+0.40 \\ -0.35}$}} &\textbf{ 0.28}\\
Log-normal: $\mu=1.0$, $\sigma=0.4$ & 20 &
$60${\raisebox{0.5ex}{\tiny$\substack{+27 \\ -29}$}} & $23${\raisebox{0.5ex}{\tiny$\substack{+11 \\ -11}$}} & 1.49 \\
Uniform on [0.01,1000] & 45 & 
$2.3${\raisebox{0.5ex}{\tiny$\substack{+1.7 \\ -0.92}$}} & $2.6${\raisebox{0.5ex}{\tiny$\substack{+1.6 \\ -0.91}$}} & 0.53\\
Log-uniform on [0.01,100] & 45 &
$1.5${\raisebox{0.5ex}{\tiny$\substack{+0.95 \\ -0.56}$}} & $1.8${\raisebox{0.5ex}{\tiny$\substack{+0.94 \\ -0.59}$}} & 0.38 \\
Truncated Gaussian: $\mu=0.69$, $\sigma=1.0$ & 45 &
$1.4${\raisebox{0.5ex}{\tiny$\substack{+0.35 \\ -0.32}$}} & $1.2${\raisebox{0.5ex}{\tiny$\substack{+0.30 \\ -0.25}$}} & \textbf{0.22} \\
Log-normal: $\mu=1.0$, $\sigma=0.4$ & 45 &
$8.2${\raisebox{0.5ex}{\tiny$\substack{+4.6 \\ -2.5}$}} & $7.6${\raisebox{0.5ex}{\tiny$\substack{+3.8 \\ -2.3}$}} & 1.01 \\
\enddata
\end{deluxetable}

\begin{figure*}
    \centering
    \includegraphics[width=\textwidth]{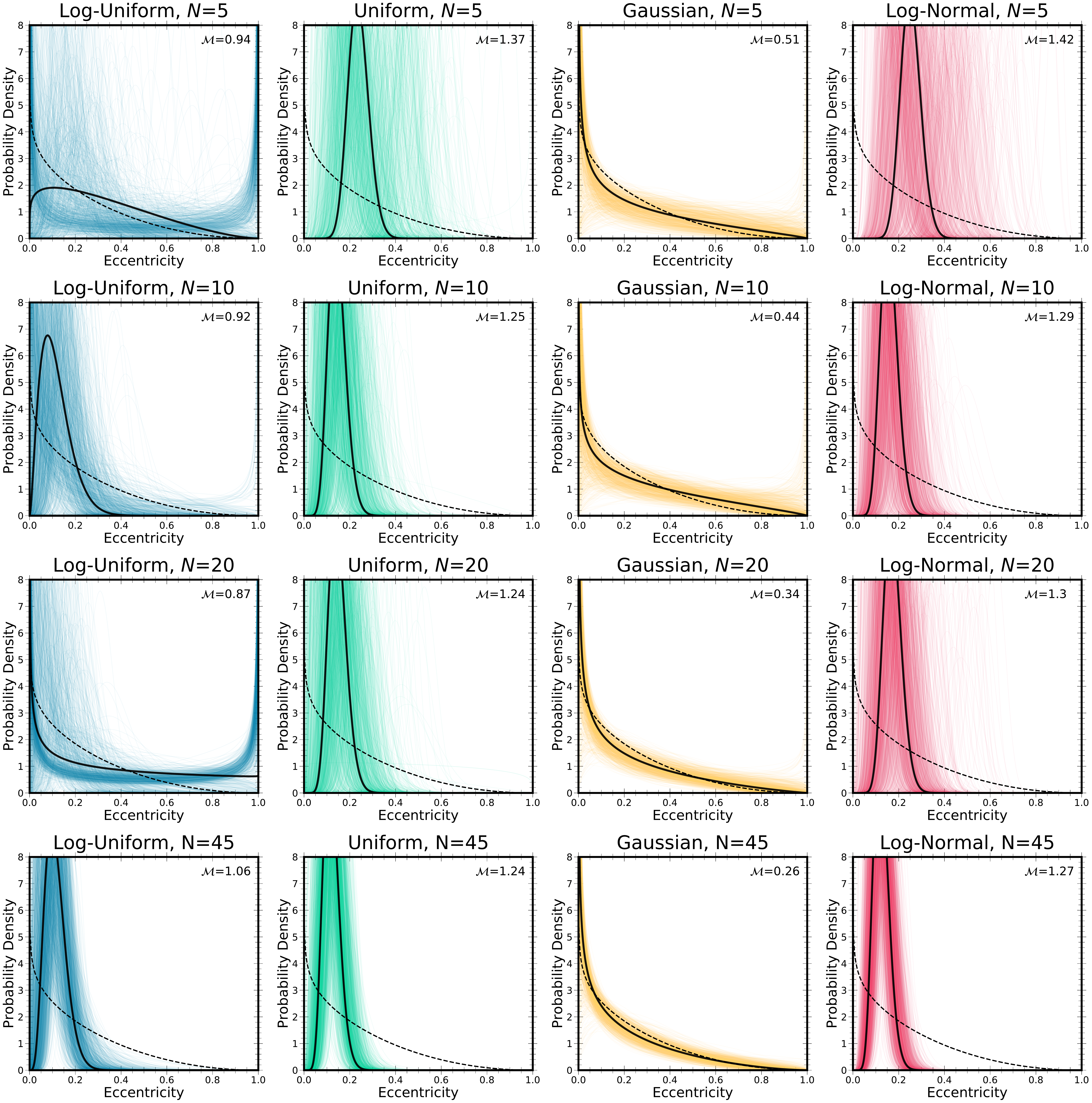}
    \caption{Same as Figure \ref{fig:uniform_cosinc}, but for the case of an underlying RV exoplanet distribution. We once again observe that the Gaussian hyperprior consistently performs the best at each sample size, followed by the log-uniform, uniform, and log-normal hyperpriors. Note that even for large samples, the log-uniform, uniform, and log-normal hyperpriors introduce a narrow peak in the posterior distributions.}
    \label{fig:wj_fixedinc}

\end{figure*}

Forward modeling an underlying RV exoplanet eccentricity distribution (Figure \ref{fig:wj_fixedinc}) yields similar results: Once again, the Gaussian hyperprior most consistently infers the correct underlying distribution, followed by the log-uniform, uniform, and log-normal hyperpriors. The corresponding values of $\alpha$, $\beta$, and $\mathcal{M}$ for each sub-sample fit are summarised in Table \ref{table:forward_wj}.

The major takeaways from our forward modeling exploration are as follows: 

\begin{enumerate}
    \item We qualitatively recover both the uniform and RV exoplanet input eccentricity distributions under realistic observational datasets of imaged planets by imposing a truncated Gaussian hyperprior  with $\mu=0.69$ and $\sigma=1.0$ on both $\alpha$ and $\beta$. 
    
    \item For both underlying distributions, and for each sample size tested, the truncated Gaussian hyperprior recovers the most accurate population-level distributions, an indication that it is robust to biases that arise when fitting the orbits of directly imaged systems with low orbital coverage. We interpret this as evidence that the truncated Gaussian hyperprior with $\mu=0.69$ and $\sigma=1.0$ is well-suited for use in efforts to infer eccentricity distributions using HBM. 
    
    \item Using a uniform hyperprior imposes a non-physical peak in the recovered distributions, similar to the results observed for the imaged giant planet population in BBN20. 

\end{enumerate}

\begin{deluxetable}{ccccc}
\tabletypesize{\footnotesize}
\tablecolumns{4}
\tablewidth{\textwidth}
\tablecaption{Analogous to Table \ref{table:forward_uniform}, but for an underlying RV exoplanet eccentricity distribution ($\alpha=0.87$, $\beta=3.03$). See Figure \ref{fig:wj_fixedinc} for a visualization of these distributions. {\fontfamily{pcr}\selectfont
} \label{table:forward_wj}}
\tablehead{
 \colhead{Hyperprior} & \colhead{$N$} & \colhead{$\alpha$} & \colhead{$\beta$} &
 \colhead{$\mathcal{M}$}
}
\startdata
Uniform on [0.01,100] & 5 &  $22${\raisebox{0.5ex}{\tiny$\substack{+25 \\ -11}$}} & $69${\raisebox{0.5ex}{\tiny$\substack{+22 \\ -31}$}} & 1.36 \\
Log-uniform on [0.01,100] & 5 & $1.2${\raisebox{0.5ex}{\tiny$\substack{+7.1 \\ -0.9}$}} & $2.6${\raisebox{0.5ex}{\tiny$\substack{+29.6 \\ -2.2}$}} & 0.95 \\
Truncated Gaussian: $\mu$=0.69, $\sigma$=1.0 & 5 & $0.8${\raisebox{0.5ex}{\tiny$\substack{+0.4 \\ -0.3}$}} & $1.3${\raisebox{0.5ex}{\tiny$\substack{+0.4 \\ -0.4}$}} & 0.67 \\
Log-normal: $\mu$=1.0, $\sigma$=0.4 & 5  & $25${\raisebox{0.5ex}{\tiny$\substack{+27 \\ -11}$}} & $73${\raisebox{0.5ex}{\tiny$\substack{+20 \\ -27}$}} & 1.43 \\
Uniform on [0.01,1000] & 10 &
$12.2${\raisebox{0.5ex}{\tiny$\substack{+7.7 \\ -5.9}$}} & $72${\raisebox{0.5ex}{\tiny$\substack{+20 \\ -30}$}} & 1.24 \\
Log-uniform on [0.01,100] & 10 &
$2.7${\raisebox{0.5ex}{\tiny$\substack{+6.7 \\ -1.9}$}} & $20${\raisebox{0.5ex}{\tiny$\substack{+41 \\ -16}$}} & 0.9 \\
Truncated Gaussian: $\mu$=0.69, $\sigma$=1.0 & 10 &
$0.6${\raisebox{0.5ex}{\tiny$\substack{+0.3 \\ -0.2}$}} & $1.4${\raisebox{0.5ex}{\tiny$\substack{+0.4 \\ -0.4}$}} & 0.5\\
Log-normal: $\mu$=1.0, $\sigma$=0.4 & 10 &
$15${\raisebox{0.5ex}{\tiny$\substack{+7.5 \\ -5.1}$}} & $78${\raisebox{0.5ex}{\tiny$\substack{+16 \\ -24}$}} & 1.29 \\
Uniform on [0.01,1000] & 20 & 
$12${\raisebox{0.5ex}{\tiny$\substack{+9.4 \\ -7.2}$}} & $69${\raisebox{0.5ex}{\tiny$\substack{+22 \\ -32}$}} & 1.24 \\
Log-uniform on [0.01,100] & 20 &
$0.59${\raisebox{0.5ex}{\tiny$\substack{+2.5 \\ -0.29}$}} & $1.0${\raisebox{0.5ex}{\tiny$\substack{+26 \\ -0.64}$}} & 0.87 \\
Truncated Gaussian: $\mu$=0.69, $\sigma$=1.0 & 20 &
$0.60${\raisebox{0.5ex}{\tiny$\substack{+0.29 \\ -0.21}$}} & $1.23${\raisebox{0.5ex}{\tiny$\substack{+0.49 \\ -0.48}$}} & 0.55 \\
Log-normal: $\mu$=1.0, $\sigma$=0.4 & 20 &
$16${\raisebox{0.5ex}{\tiny$\substack{+8.5 \\ -6.3}$}} & $77${\raisebox{0.5ex}{\tiny$\substack{+17 \\ -25}$}} & 1.3 \\
Uniform on [0.01,1000] & 45 & 
$10${\raisebox{0.5ex}{\tiny$\substack{+3.7 \\ -3.9}$}} & $76${\raisebox{0.5ex}{\tiny$\substack{+18 \\ -27}$}} & 1.23 \\
Log-uniform on [0.01,100] & 45 &
$5.4${\raisebox{0.5ex}{\tiny$\substack{+5.2 \\ -3.5}$}} & $76${\raisebox{0.5ex}{\tiny$\substack{+18 \\ -27}$}} & 1.06 \\
Truncated Gaussian: $\mu$=0.69, $\sigma$=1.0 & 45 &
$0.60${\raisebox{0.5ex}{\tiny$\substack{+0.20 \\ -0.15}$}} & $1.7${\raisebox{0.5ex}{\tiny$\substack{+0.43 \\ -0.43}$}} & 0.35 \\
Log-normal: $\mu$=1.0, $\sigma$=0.4 & 45 &
$12${\raisebox{0.5ex}{\tiny$\substack{+3.3 \\ -3.2}$}} & $82${\raisebox{0.5ex}{\tiny$\substack{+13 \\ -21}$}} & 1.27 \\
\enddata
\end{deluxetable}

\section{Re-analysing the imaged substellar companion sample} \label{re-analysis}

We now use \texttt{ePop!}'s HBM functionality to fit for the population-level eccentricity distributions underlying the sets of imaged companions analysed by BBN20 using lessons learned from our experiments. This sample contains a total of 27 substellar companions separated by a  mass threshold of 15 $M_\mathrm{Jup}$ into subsamples of 9 giant planets and 18 brown dwarfs. To facilitate a direct comparison and highlight the impact of hyperprior choice, we conduct our analysis on this sample of substellar companions using the exact eccentricity posteriors as BBN20. See Table 5 in BBN20 for details about the systems contained in this sample, as well as of the origins of the eccentricity posteriors we make use of in this study. 

Following the approach of previous sections, we perform fits using the following three hyperpriors on ($\alpha$, $\beta$): truncated Gaussian with $\mu=0.69$ and $\sigma=1$ , uniform on [0.01,1000], and log-uniform on [0.01,100]. The findings of Section \ref{forward_modelling_section} imply that the truncated Gaussian hyperprior is able to recover qualitatively accurate underlying eccentricity posteriors for small samples, at least in the cases of underlying uniform and RV exoplanet eccentricity distributions. Motivated by these results, we argue that the eccentricity distributions recovered by the truncated Gaussian hyperprior are more likely to be accurate, as compared to those recovered by the uniform and log-uniform hyperpriors. 

\begin{figure*}
    \includegraphics[width=\textwidth]{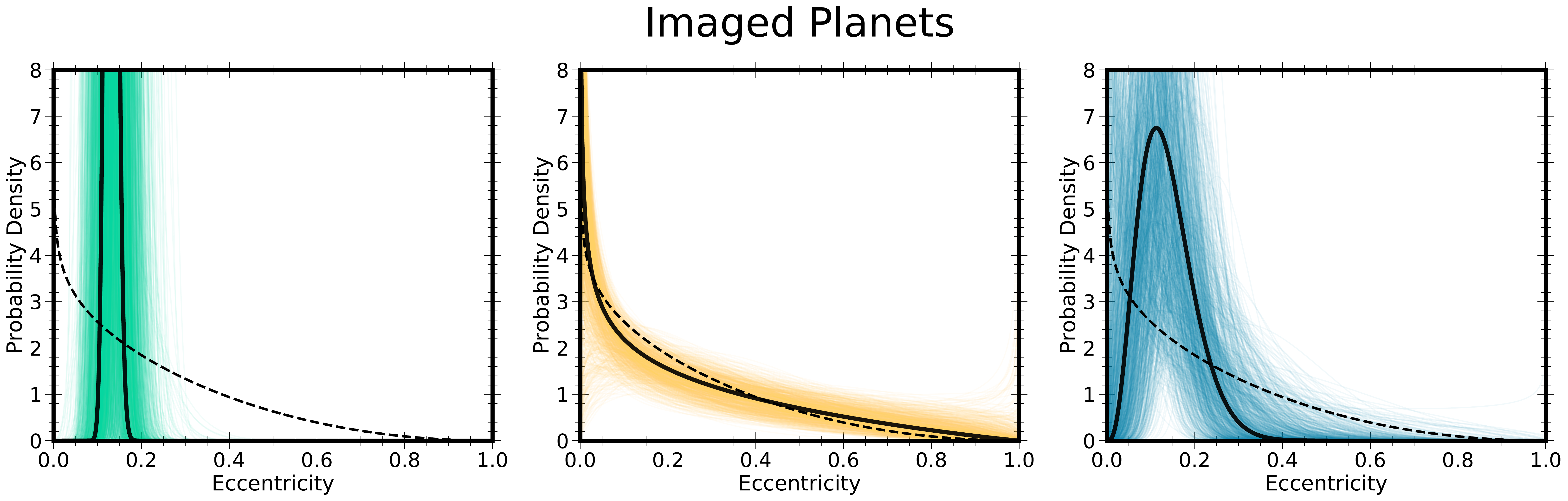}
    \caption{Results of applying HBM to the eccentricity posteriors for the sample of 9 imaged planets from BBN20. \textit{Left panel}: Eccentricity distributions inferred using a uniform hyperprior on ($\alpha$, $\beta$) with bounds [0.01,1000]. These results mirror those from BBN20. \textit{Middle panel}: Distributions recovered by imposing a truncated Gaussian hyperprior with $\mu=0.69$ and $\sigma$=1.0. Shown for reference (dashed line) is the RV exoplanet eccentricity distribution. Based on the results of the hyperprior tests in Sections \ref{Gaussian} and \ref{forward_modelling_section}, we interpret this as evidence for a similarity between the eccentricity distributions of widely separated imaged giant planets and close-in RV exoplanets.
    \textit{Right panel}: Results obtained by imposing a log-uniform hyperprior with bounds [0.01,100]. We note the similarity between the distributions recovered for this sample ($N$=9) and the distributions recovered by the log-uniform prior for the $N$=10 case in Figure \ref{fig:wj_fixedinc}.
    \label{fig:gp_reanalysis}}
\end{figure*}

For the imaged planet sample, using the truncated Gaussian hyperprior yields $\alpha=0.7${\raisebox{0.5ex}{\tiny$\substack{+0.4 \\ -0.3}$}}
and $\beta=2.3${\raisebox{0.5ex}{\tiny$\substack{+0.8 \\ -0.7}$}}, a range of hyperparameters corresponding to the family of distributions shown in the middle panel of Figure \ref{fig:gp_reanalysis}. Notably, this range encompasses the RV exoplanet eccentricity distribution ($\alpha$=0.87, $\beta$=3.03, from \citealt{kipping}). This suggests that the eccentricity distribution underlying the population of imaged giant planets is consistent with the eccentricity distribution of RV exoplanets that reside much closer to their host stars. A similar result was found by BBN20 using a mass ratio threshold.

\begin{deluxetable*}{cccc}
\tabletypesize{\footnotesize}
\tablecolumns{2}
\tablewidth{\textwidth}
\tablecaption{Hyperparameter ranges corresponding to the recovered eccentricity distributions from our HBM fits to the imaged planet and brown dwarf subsamples. The distributions themselves are shown in Figures  \ref{fig:gp_reanalysis} and \ref{fig:bd_reanalysis}.} \label{table:reanalysis_results}
\tablehead{
 \colhead{Sample} & \colhead{Hyperprior} & \colhead{$\alpha$} & \colhead{$\beta$}
}
\startdata
Giant Planets & Truncated Gaussian: $\mu=0.69$, $\sigma=1.0$ & $0.7^{+0.4}_{-0.3}$ & $2.3^{+0.8}_{-0.7}$ \\
Giant Planets & Log-Uniform on [0.01,100] & $4.2${\raisebox{0.5ex}{\tiny$\substack{+6.2 \\ -2.8}$}}
& $26${\raisebox{0.5ex}{\tiny$\substack{+40 \\ -18}$}} \\
Giant Planets & Uniform on [0.01,1000] & $106${\raisebox{0.5ex}{\tiny$\substack{+45 \\ -48}$}}
& $699${\raisebox{0.5ex}{\tiny$\substack{+216 \\ -314}$}} \\
Brown Dwarfs & Truncated Gaussian:$\mu=0.69, \sigma=1.0$ & $1.6${\raisebox{0.5ex}{\tiny$\substack{+0.6 \\ -0.5}$}}
& $1.2${\raisebox{0.5ex}{\tiny$\substack{+0.5 \\ -0.4}$}} \\
Brown Dwarfs & Log-Uniform on [0.01,100] & $1.5${\raisebox{0.5ex}{\tiny$\substack{+1.1 \\ -0.6}$}}
& $1.2${\raisebox{0.5ex}{\tiny$\substack{+0.7 \\ -0.4}$}} \\
Brown Dwarfs & Uniform on [0.01,100] & $2.5${\raisebox{0.5ex}{\tiny$\substack{+1.7 \\ -1.1}$}}
& $1.8${\raisebox{0.5ex}{\tiny$\substack{+1.1 \\ -0.7}$}} \\
\enddata
\end{deluxetable*}
 
\begin{figure*}
    \includegraphics[width=\textwidth]{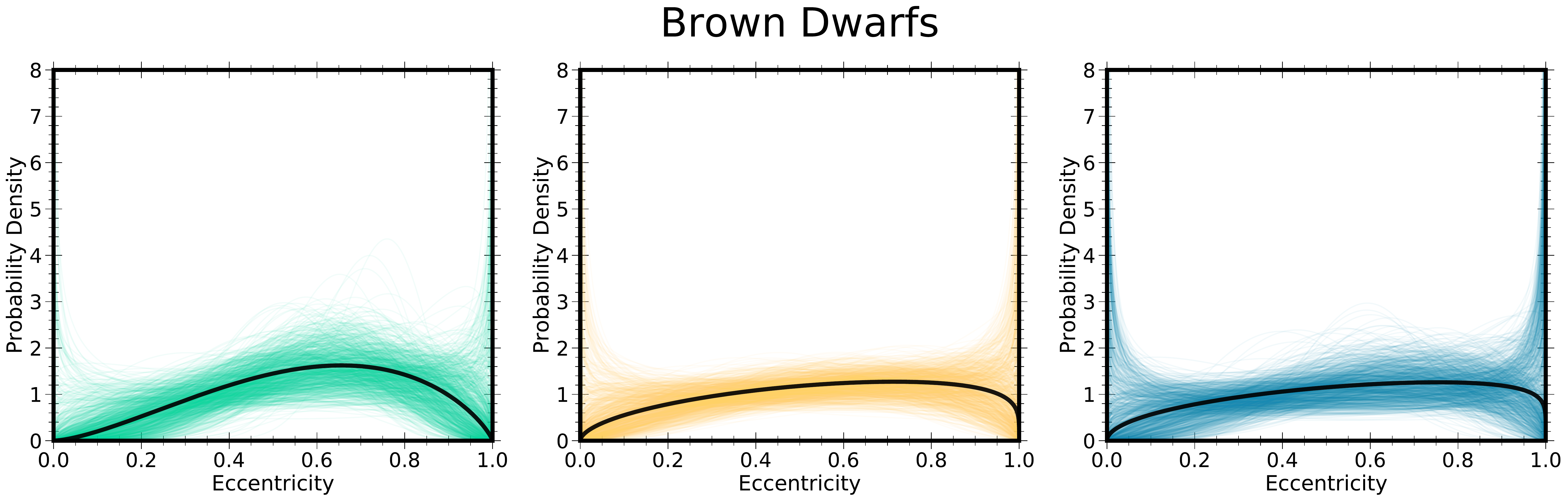}
    \caption{Same as Figure \ref{fig:gp_reanalysis}, but for the sample of 18 brown dwarfs from BBN20. The hyperprior does not exert a significant influence on the inferred population-level eccentricity distributions for this sample. }
    \label{fig:bd_reanalysis}
\end{figure*}

Using a log-uniform prior results in significantly different recovered distributions: the hyperparameter ranges in this case shift to $\alpha=4.2${\raisebox{0.5ex}{\tiny$\substack{+6.2 \\ -2.8}$}} and $ \beta=26${\raisebox{0.5ex}{\tiny$\substack{+40 \\ -18}$}}). Interestingly, there is a striking resemblance between these results for the imaged planet sample (right panel of Figure \ref{fig:gp_reanalysis}, $N$=9) and the distributions recovered for the $N$=10 sample of eccentricity posteriors for forward modeled systems drawn from an assumed underlying RV exoplanet eccentricity distribution (Figure \ref{fig:wj_fixedinc})---another possible indication that imaged planets have a similar eccentricity distribution to RV exoplanets. Finally, applying a uniform hyperprior yields $\alpha=106${\raisebox{0.5ex}{\tiny$\substack{+45 \\ -48}$}},  $ \beta=699${\raisebox{0.5ex}{\tiny$\substack{+216 \\ -314}$}}. This corresponds to a range of narrow distributions (left panel of Figure \ref{fig:gp_reanalysis}) with pronounced peaks at $e\sim0.15$---mirroring the results from BBN 2020. The narrowly peaked nature of this family of distributions is likely a consequence of biases imparted by the uniform hyperprior (Figure \ref{beta_hyperprior}).

For the brown dwarf subsample, we find that the choice of hyperprior does not exert a significant impact on the inferred eccentricity distributions. As can be seen in Figure \ref{fig:bd_reanalysis}, the hyperpriors we test recover similar eccentricity distributions. The results of each fit are consistent with those of BBN20, who (using a uniform hyperprior) found $\alpha=2.3$ and $\beta=1.7$ for the best-fit hyperparameters for the brown dwarf subsample. The decreased influence of hyperprior choice in this case can be explained by its larger sample size ($N$=18) relative to the imaged planet subsample ($N$=9) as well as the presence of brown dwarfs with tightly constrained non-zero and non-overlapping eccentricity posteriors, which disfavors the narrowly peaked eccentricity distributions preferred by the uniform and log-uniform hyperpriors. In contrast, the individual eccentricity posteriors for the imaged planets in the sample tend to be much broader, so the choice of hyperprior more readily influences the inferred population level distributions. We note, however, that our results show a clear difference between the eccentricity distributions of the imaged planet and brown dwarf samples---a distinction that is robust to hyperprior choice.

\section{Conclusion}

In this study, we performed a systematic exploration of the impact that the choice of hyperprior has on the population-level eccentricity distributions recovered using HBM for a model parameterized by a Beta distribution. Our key finding is that imposing a uniform prior on the hyperparameters ($\alpha$, $\beta$)---a choice often made in previous studies attempting such population level analysis---imparts a significant bias onto the resulting posteriors when the sample size is small or when individual eccentricity constraints are large. This can be explained by the fact that a uniform prior in ($\alpha$, $\beta$) space corresponds to a narrowly peaked family of Beta distributions in eccentricity space. Moreover, the wider the hyperparameter range, the narrower the eccentricity hyperprior distributions become. As such, we argue that a uniform hyperprior on ($\alpha$, $\beta$) does not accurately represent our prior expectation for population-level eccentricity distributions and should not be used as a default choice for HBM with the Beta Distribution.

Instead, a truncated Gaussian hyperprior (with $\mu=0.69, \sigma=1.0$) on the Beta distribution hyperparameters appears to be a much more suitable hyperprior. The family of distributions it produces cover a wide range of morphologies, including the shapes predicted by different planet formation theories. We thus recommend the truncated Gaussian hyperprior with ($\mu, \sigma$) =(0.69, 1.0) for use as a weakly informative hyperprior when performing HBM using the Beta distribution. 

We performed a series of forward modeling experiments to explore how well we can recover a known underlying eccentricity distribution from a sample of substellar companions. Two assumed underlying distributions were explored, the RV exoplanet distribution from \citet{kipping}, and the uniform eccentricity distribution, using four qualitatively distinct hyperpriors. In both cases, and for all sample sizes tested, we found that a truncated Gaussian prior with $\mu=0.69$ and $\sigma=1.0$ performed the best and was able to most consistently recover the input distributions. 

Applying these findings to the analysis of real data for imaged substellar companions, we first confirm the key finding of BBN20: imaged giant planets and brown dwarfs have different eccentricity distributions, a distinction that is robust to choice of hyperprior and points to separate formational pathways for the two classes of objects. This echoes the results of \cite{nielsen}, who found different distributions in mass and semi-major axis for imaged giant planets and brown dwarfs. We calculate new results for the population-level eccentricity distribution of widely separated imaged planets and find that its underlying eccentricity distribution is similar to the eccentricity distribution of RV exoplanets from \citet{kipping}. Given the small size of the current imaged planet sample, there is a need for additional discoveries and continued monitoring of known companions to firmly establish the exact degree of similarity between the eccentricities of these two classes of planets. However, even at this stage, these findings raise an interesting question regarding planet formation: are the mechanisms involved in the formation and subsequent dynamical evolution of close-in planets in the RV exoplanet sample related to those responsible for forming gas giants imaged at wide separations observed using direct imaging? Or is this resemblance coincidental? BBN20 recovered a distribution similar to RV exoplanets for imaged giant planets when the observational sample of substellar companions was split by mass ratio rather than mass, indicating the sensitivity of the population-level eccentricity posterior to the exact make up of the sample.

There are many possible extensions to the method we have developed in this study that may help elucidate our understanding of the formation of widely separated giant planets. The forward modeling can be made even more realistic by incorporating a mixture of well-constrained and less-constrained posteriors to better mimic the status of observational samples today. Furthermore, simulating data with multiple observation types (such as radial velocities and sky-plane astrometric accelerations from Gaia/Hipparcos) in addition to astrometry from high-contrast imaging would make the method more realistic. In the future, with more systems in hand, eccentricity distributions could be studied as functions of other quantities of interest such as planet mass, stellar metallicity, and separation, which will provide a more granular understanding of the factors affecting planet formation and dynamical evolution. Furthermore, leveraging such a sample will enable expansion beyond the Beta distribution and allow for detailed model comparisons that will help determine the most reliable model choices for inferring eccentricity distributions with HBM. Together with advances in individual orbit characterisation made possible by micro-arcsecond astrometry from instruments such as GRAVITY (\citealt{,micro}; \citealt{gravity}, \citealt{hr8799}), hierarchical inference of planet population eccentricities will establish a more complete view of the planet formation process.

\begin{acknowledgements}
We thank the anonymous referee for constructive suggestions that improved the quality of this manuscript. V.N., S.B, and J.J.W. acknowledge support from the Heising-Simons Foundation, including grant 2019-1698. B.P.B. acknowledges support from the National Science Foundation grant AST-1909209, NASA Exoplanet Research Program grant 20-XRP20$\_$2-0119, and the Alfred P. Sloan Foundation. For the purpose of open access, the author has applied a Creative Commons Attribution (CC BY) license to any Author Accepted Manuscript version arising from this submission.

\end{acknowledgements}

\bibliography{eccentricities}

\begin{thebibliography}{}
\expandafter\ifx\csname natexlab\endcsname\relax\def\natexlab#1{#1}\fi
\providecommand{\url}[1]{\href{#1}{#1}}
\providecommand{\dodoi}[1]{doi:~\href{http://doi.org/#1}{\nolinkurl{#1}}}
\providecommand{\doeprint}[1]{\href{http://ascl.net/#1}{\nolinkurl{http://ascl.net/#1}}}
\providecommand{\doarXiv}[1]{\href{https://arxiv.org/abs/#1}{\nolinkurl{https://arxiv.org/abs/#1}}}

\bibitem[{{Armitage}(2013)}]{armitagebook}
{Armitage}, P.~J. 2013, {Astrophysics of Planet Formation}

\bibitem[{{Bate}(2012)}]{bate}
{Bate}, M.~R. 2012, \mnras, 419, 3115, \dodoi{10.1111/j.1365-2966.2011.19955.x}

\bibitem[{{Blunt} {et~al.}(2017){Blunt}, {Nielsen}, {De Rosa}, {Konopacky},
  {Ryan}, {Wang}, {Pueyo}, {Rameau}, {Marois}, {Marchis}, {Macintosh},
  {Graham}, {Duch{\^e}ne}, \& {Schneider}}]{OFTI}
{Blunt}, S., {Nielsen}, E.~L., {De Rosa}, R.~J., {et~al.} 2017, \aj, 153, 229,
  \dodoi{10.3847/1538-3881/aa6930}

\bibitem[{{Blunt} {et~al.}(2020){Blunt}, {Wang}, {Angelo}, {Ngo}, {Cody}, {De
  Rosa}, {Graham}, {Hirsch}, {Nagpal}, {Nielsen}, {Pearce}, {Rice}, \&
  {Tejada}}]{orbitize}
{Blunt}, S., {Wang}, J.~J., {Angelo}, I., {et~al.} 2020, \aj, 159, 89,
  \dodoi{10.3847/1538-3881/ab6663}

\bibitem[{{Bowler}(2016)}]{bowlerreview}
{Bowler}, B.~P. 2016, \pasp, 128, 102001,
  \dodoi{10.1088/1538-3873/128/968/102001}

\bibitem[{{Bowler} {et~al.}(2020){Bowler}, {Blunt}, \& {Nielsen}}]{bowler:2020}
{Bowler}, B.~P., {Blunt}, S.~C., \& {Nielsen}, E.~L. 2020, \aj, 159, 63,
  \dodoi{10.3847/1538-3881/ab5b11}

\bibitem[{{Chauvin} {et~al.}(2012){Chauvin}, {Lagrange}, {Beust}, {Bonnefoy},
  {Boccaletti}, {Apai}, {Allard}, {Ehrenreich}, {Girard}, {Mouillet}, \&
  {Rouan}}]{chauvin}
{Chauvin}, G., {Lagrange}, A.~M., {Beust}, H., {et~al.} 2012, \aap, 542, A41,
  \dodoi{10.1051/0004-6361/201118346}

\bibitem[{{Dawson} \& {Murray-Clay}(2013)}]{dawson_murrayclay}
{Dawson}, R.~I., \& {Murray-Clay}, R.~A. 2013, \apjl, 767, L24,
  \dodoi{10.1088/2041-8205/767/2/L24}

\bibitem[{{Dong} {et~al.}(2021){Dong}, {Huang}, {Dawson}, {Foreman-Mackey},
  {Collins}, {Quinn}, {Lissauer}, {Beatty}, {Quarles}, {Sha}, {Shporer}, {Guo},
  {Kane}, {Abe}, {Barkaoui}, {Benkhaldoun}, {Brahm}, {Bouchy}, {Carmichael},
  {Collins}, {Conti}, {Crouzet}, {Dransfield}, {Evans}, {Gan}, {Ghachoui},
  {Gillon}, {Grieves}, {Guillot}, {Hellier}, {Jehin}, {Jensen}, {Jord{\'a}n},
  {Kamler}, {Kielkopf}, {M{\'e}karnia}, {Nielsen}, {Pozuelos}, {Radford},
  {Schmider}, {Schwarz}, {Stockdale}, {Tan}, {Timmermans}, {Triaud}, {Wang},
  {Ricker}, {Vanderspek}, {Latham}, {Seager}, {Winn}, {Jenkins}, {Mireles},
  {Yahalomi}, {Morgan}, {Vezie}, {Quintana}, {Rose}, {Smith}, \&
  {Shiao}}]{tess_beta}
{Dong}, J., {Huang}, C.~X., {Dawson}, R.~I., {et~al.} 2021, \apjs, 255, 6,
  \dodoi{10.3847/1538-4365/abf73c}

\bibitem[{{Duffell} \& {Chiang}(2015)}]{chiang_duffell}
{Duffell}, P.~C., \& {Chiang}, E. 2015, \apj, 812, 94,
  \dodoi{10.1088/0004-637X/812/2/94}

\bibitem[{{Ferrer-Ch{\'a}vez} {et~al.}(2021){Ferrer-Ch{\'a}vez}, {Wang}, \&
  {Blunt}}]{rodrigo}
{Ferrer-Ch{\'a}vez}, R., {Wang}, J.~J., \& {Blunt}, S. 2021, \aj, 161, 241,
  \dodoi{10.3847/1538-3881/abf0a8}

\bibitem[{{Foreman-Mackey} {et~al.}(2013){Foreman-Mackey}, {Hogg}, {Lang}, \&
  {Goodman}}]{emcee}
{Foreman-Mackey}, D., {Hogg}, D.~W., {Lang}, D., \& {Goodman}, J. 2013, \pasp,
  125, 306, \dodoi{10.1086/670067}

\bibitem[{{Goldreich} \& {Sari}(2003)}]{goldreich_sari}
{Goldreich}, P., \& {Sari}, R. 2003, \apj, 585, 1024, \dodoi{10.1086/346202}

\bibitem[{{Gravity Collaboration} {et~al.}(2017){Gravity Collaboration},
  {Abuter}, {Accardo}, {Amorim}, {Anugu}, {{\'A}vila}, {Azouaoui}, {Benisty},
  {Berger}, {Blind}, {Bonnet}, {Bourget}, {Brandner}, {Brast}, {Buron},
  {Burtscher}, {Cassaing}, {Chapron}, {Choquet}, {Cl{\'e}net}, {Collin},
  {Coud{\'e} Du Foresto}, {de Wit}, {de Zeeuw}, {Deen},
  {Delplancke-Str{\"o}bele}, {Dembet}, {Derie}, {Dexter}, {Duvert}, {Ebert},
  {Eckart}, {Eisenhauer}, {Esselborn}, {F{\'e}dou}, {Finger}, {Garcia}, {Garcia
  Dabo}, {Garcia Lopez}, {Gendron}, {Genzel}, {Gillessen}, {Gonte}, {Gordo},
  {Grould}, {Gr{\"o}zinger}, {Guieu}, {Haguenauer}, {Hans}, {Haubois}, {Haug},
  {Haussmann}, {Henning}, {Hippler}, {Horrobin}, {Huber}, {Hubert}, {Hubin},
  {Hummel}, {Jakob}, {Janssen}, {Jochum}, {Jocou}, {Kaufer}, {Kellner},
  {Kendrew}, {Kern}, {Kervella}, {Kiekebusch}, {Klein}, {Kok}, {Kolb}, {Kulas},
  {Lacour}, {Lapeyr{\`e}re}, {Lazareff}, {Le Bouquin}, {L{\`e}na}, {Lenzen},
  {L{\'e}v{\^e}que}, {Lippa}, {Magnard}, {Mehrgan}, {Mellein}, {M{\'e}rand},
  {Moreno-Ventas}, {Moulin}, {M{\"u}ller}, {M{\"u}ller}, {Neumann}, {Oberti},
  {Ott}, {Pallanca}, {Panduro}, {Pasquini}, {Paumard}, {Percheron}, {Perraut},
  {Perrin}, {Pfl{\"u}ger}, {Pfuhl}, {Phan Duc}, {Plewa}, {Popovic}, {Rabien},
  {Ram{\'\i}rez}, {Ramos}, {Rau}, {Riquelme}, {Rohloff}, {Rousset},
  {Sanchez-Bermudez}, {Scheithauer}, {Sch{\"o}ller}, {Schuhler}, {Spyromilio},
  {Straubmeier}, {Sturm}, {Suarez}, {Tristram}, {Ventura}, {Vincent},
  {Waisberg}, {Wank}, {Weber}, {Wieprecht}, {Wiest}, {Wiezorrek}, {Wittkowski},
  {Woillez}, {Wolff}, {Yazici}, {Ziegler}, \& {Zins}}]{gravity}
{Gravity Collaboration}, {Abuter}, R., {Accardo}, M., {et~al.} 2017, \aap, 602,
  A94, \dodoi{10.1051/0004-6361/201730838}

\bibitem[{{Gravity Collaboration} {et~al.}(2019){Gravity Collaboration},
  {Lacour}, {Nowak}, {Wang}, {Pfuhl}, {Eisenhauer}, {Abuter}, {Amorim},
  {Anugu}, {Benisty}, {Berger}, {Beust}, {Blind}, {Bonnefoy}, {Bonnet},
  {Bourget}, {Brandner}, {Buron}, {Collin}, {Charnay}, {Chapron}, {Cl{\'e}net},
  {Coud{\'e} Du Foresto}, {de Zeeuw}, {Deen}, {Dembet}, {Dexter}, {Duvert},
  {Eckart}, {F{\"o}rster Schreiber}, {F{\'e}dou}, {Garcia}, {Garcia Lopez},
  {Gao}, {Gendron}, {Genzel}, {Gillessen}, {Gordo}, {Greenbaum}, {Habibi},
  {Haubois}, {Hau{\ss}mann}, {Henning}, {Hippler}, {Horrobin}, {Hubert},
  {Jimenez Rosales}, {Jocou}, {Kendrew}, {Kervella}, {Kolb}, {Lagrange},
  {Lapeyr{\`e}re}, {Le Bouquin}, {L{\'e}na}, {Lippa}, {Lenzen}, {Maire},
  {Molli{\`e}re}, {Ott}, {Paumard}, {Perraut}, {Perrin}, {Pueyo}, {Rabien},
  {Ram{\'\i}rez}, {Rau}, {Rodr{\'\i}guez-Coira}, {Rousset}, {Sanchez-Bermudez},
  {Scheithauer}, {Schuhler}, {Straub}, {Straubmeier}, {Sturm}, {Tacconi},
  {Vincent}, {van Dishoeck}, {von Fellenberg}, {Wank}, {Waisberg}, {Widmann},
  {Wieprecht}, {Wiest}, {Wiezorrek}, {Woillez}, {Yazici}, {Ziegler}, \&
  {Zins}}]{hr8799}
{Gravity Collaboration}, {Lacour}, S., {Nowak}, M., {et~al.} 2019, \aap, 623,
  L11, \dodoi{10.1051/0004-6361/201935253}

\bibitem[{{Hogg} {et~al.}(2010){Hogg}, {Myers}, \& {Bovy}}]{Hogg}
{Hogg}, D.~W., {Myers}, A.~D., \& {Bovy}, J. 2010, \apj, 725, 2166,
  \dodoi{10.1088/0004-637X/725/2/2166}

\bibitem[{{Juri{\'c}} \& {Tremaine}(2008)}]{juric_tremaine}
{Juri{\'c}}, M., \& {Tremaine}, S. 2008, \apj, 686, 603, \dodoi{10.1086/590047}

\bibitem[{{Kipping}(2013)}]{kipping}
{Kipping}, D.~M. 2013, \mnras, 434, L51, \dodoi{10.1093/mnrasl/slt075}

\bibitem[{{Konopacky} {et~al.}(2016){Konopacky}, {Marois}, {Macintosh},
  {Galicher}, {Barman}, {Metchev}, \& {Zuckerman}}]{konopacky}
{Konopacky}, Q.~M., {Marois}, C., {Macintosh}, B.~A., {et~al.} 2016, \aj, 152,
  28, \dodoi{10.3847/0004-6256/152/2/28}

\bibitem[{{Lacour} {et~al.}(2014){Lacour}, {Eisenhauer}, {Gillessen}, {Pfuhl},
  {Woillez}, {Bonnet}, {Perrin}, {Lazareff}, {Rabien}, {Lapeyr{\`e}re},
  {Cl{\'e}net}, {Kervella}, \& {Kok}}]{micro}
{Lacour}, S., {Eisenhauer}, F., {Gillessen}, S., {et~al.} 2014, \aap, 567, A75,
  \dodoi{10.1051/0004-6361/201423940}

\bibitem[{{Morbidelli}(2018)}]{morbidelli}
{Morbidelli}, A. 2018, in Handbook of Exoplanets, ed. H.~J. {Deeg} \& J.~A.
  {Belmonte}, 145, \dodoi{10.1007/978-3-319-55333-7\_145}

\bibitem[{{Nagpal} {et~al.}(2022){Nagpal}, {Blunt}, {Bolwer}, {Dupuy},
  {Nielsen}, \& {Wang}}]{ePop_code}
{Nagpal}, V., {Blunt}, S., {Bolwer}, B.~P., {et~al.} 2022, {ePop!: A Package
  for Inferring Population-level Eccentricity Distributions Using Hierarchical
  MCMC.}, 1.0,  Zenodo, \dodoi{10.5281/zenodo.7240416}

\bibitem[{{Naoz}(2016)}]{naoz}
{Naoz}, S. 2016, \araa, 54, 441, \dodoi{10.1146/annurev-astro-081915-023315}

\bibitem[{{Nielsen} {et~al.}(2019){Nielsen}, {De Rosa}, {Macintosh}, {Wang},
  {Ruffio}, {Chiang}, {Marley}, {Saumon}, {Savransky}, {Ammons}, {Bailey},
  {Barman}, {Blain}, {Bulger}, {Burrows}, {Chilcote}, {Cotten}, {Czekala},
  {Doyon}, {Duch{\^e}ne}, {Esposito}, {Fabrycky}, {Fitzgerald}, {Follette},
  {Fortney}, {Gerard}, {Goodsell}, {Graham}, {Greenbaum}, {Hibon}, {Hinkley},
  {Hirsch}, {Hom}, {Hung}, {Dawson}, {Ingraham}, {Kalas}, {Konopacky},
  {Larkin}, {Lee}, {Lin}, {Maire}, {Marchis}, {Marois}, {Metchev},
  {Millar-Blanchaer}, {Morzinski}, {Oppenheimer}, {Palmer}, {Patience},
  {Perrin}, {Poyneer}, {Pueyo}, {Rafikov}, {Rajan}, {Rameau}, {Rantakyr{\"o}},
  {Ren}, {Schneider}, {Sivaramakrishnan}, {Song}, {Soummer}, {Tallis},
  {Thomas}, {Ward-Duong}, \& {Wolff}}]{nielsen}
{Nielsen}, E.~L., {De Rosa}, R.~J., {Macintosh}, B., {et~al.} 2019, \aj, 158,
  13, \dodoi{10.3847/1538-3881/ab16e9}

\bibitem[{{Nowak} {et~al.}(2020){Nowak}, {Lacour}, {Lagrange}, {Rubini},
  {Wang}, {Stolker}, {Abuter}, {Amorim}, {Asensio-Torres}, {Baub{\"o}ck},
  {Benisty}, {Berger}, {Beust}, {Blunt}, {Boccaletti}, {Bonnefoy}, {Bonnet},
  {Brandner}, {Cantalloube}, {Charnay}, {Choquet}, {Christiaens}, {Cl{\'e}net},
  {Coud{\'e} Du Foresto}, {Cridland}, {de Zeeuw}, {Dembet}, {Dexter},
  {Drescher}, {Duvert}, {Eckart}, {Eisenhauer}, {Gao}, {Garcia}, {Garcia
  Lopez}, {Gardner}, {Gendron}, {Genzel}, {Gillessen}, {Girard}, {Grandjean},
  {Haubois}, {Hei{\ss}el}, {Henning}, {Hinkley}, {Hippler}, {Horrobin},
  {Houll{\'e}}, {Hubert}, {Jim{\'e}nez-Rosales}, {Jocou}, {Kammerer},
  {Kervella}, {Keppler}, {Kreidberg}, {Kulikauskas}, {Lapeyr{\`e}re}, {Le
  Bouquin}, {L{\'e}na}, {M{\'e}rand}, {Maire}, {Molli{\`e}re}, {Monnier},
  {Mouillet}, {M{\"u}ller}, {Nasedkin}, {Ott}, {Otten}, {Paumard}, {Paladini},
  {Perraut}, {Perrin}, {Pueyo}, {Pfuhl}, {Rameau}, {Rodet},
  {Rodr{\'\i}guez-Coira}, {Rousset}, {Scheithauer}, {Shangguan}, {Stadler},
  {Straub}, {Straubmeier}, {Sturm}, {Tacconi}, {van Dishoeck}, {Vigan},
  {Vincent}, {von Fellenberg}, {Ward-Duong}, {Widmann}, {Wieprecht},
  {Wiezorrek}, {Woillez}, \& {Gravity Collaboration}}]{betapic}
{Nowak}, M., {Lacour}, S., {Lagrange}, A.~M., {et~al.} 2020, \aap, 642, L2,
  \dodoi{10.1051/0004-6361/202039039}

\bibitem[{{Ogilvie}(2014)}]{ogilvie}
{Ogilvie}, G.~I. 2014, \araa, 52, 171,
  \dodoi{10.1146/annurev-astro-081913-035941}

\bibitem[{{Pearce} {et~al.}(2019){Pearce}, {Kraus}, {Dupuy}, {Ireland},
  {Rizzuto}, {Bowler}, {Birchall}, \& {Wallace}}]{logan}
{Pearce}, L.~A., {Kraus}, A.~L., {Dupuy}, T.~J., {et~al.} 2019, \aj, 157, 71,
  \dodoi{10.3847/1538-3881/aafacb}

\bibitem[{{Rasio} \& {Ford}(1996)}]{rasio_ford}
{Rasio}, F.~A., \& {Ford}, E.~B. 1996, Science, 274, 954,
  \dodoi{10.1126/science.274.5289.954}

\bibitem[{{Shabram} {et~al.}(2016){Shabram}, {Demory}, {Cisewski}, {Ford}, \&
  {Rogers}}]{shabram}
{Shabram}, M., {Demory}, B.-O., {Cisewski}, J., {Ford}, E.~B., \& {Rogers}, L.
  2016, \apj, 820, 93, \dodoi{10.3847/0004-637X/820/2/93}

\bibitem[{{Van Eylen} {et~al.}(2019){Van Eylen}, {Albrecht}, {Huang},
  {MacDonald}, {Dawson}, {Cai}, {Foreman-Mackey}, {Lundkvist}, {Silva Aguirre},
  {Snellen}, \& {Winn}}]{beta_smallplanets}
{Van Eylen}, V., {Albrecht}, S., {Huang}, X., {et~al.} 2019, \aj, 157, 61,
  \dodoi{10.3847/1538-3881/aaf22f}

\bibitem[{{Veras} {et~al.}(2009){Veras}, {Crepp}, \& {Ford}}]{veras}
{Veras}, D., {Crepp}, J.~R., \& {Ford}, E.~B. 2009, \apj, 696, 1600,
  \dodoi{10.1088/0004-637X/696/2/1600}

\end{thebibliography}
\bibliographystyle{aasjournal}

\end{document}